%% file: JMPS_cryo_2025.tex
\documentclass[3p]{elsarticle}
\usepackage{amsmath,amssymb,amsthm,amsfonts}
\usepackage{natbib}
\usepackage{tcolorbox}
\usepackage[font=footnotesize,labelfont=bf]{caption}
\usepackage{float}
\usepackage[framemethod=tikz]{mdframed}
\usepackage{todonotes}
\usepackage{bold-extra} 
\usepackage{graphicx,wrapfig, caption}
\usepackage{sectsty}
\usepackage{pgfplots}
\usepackage[mathscr]{euscript}

\usepackage[colorlinks=true, allcolors=blue]{hyperref}
\usepackage{enumitem} 
\usepackage{soul}
\usepackage{empheq}
\usepackage{sidecap}
\usepackage{xcolor}
\usepackage{multirow}
\usepackage{graphbox}
\usepackage{booktabs}
\usepackage{longtable}
\usepackage{verbatimbox}
\usepackage{array}
\usepackage{fancyhdr}
\usepackage{booktabs}
\usepackage{tabularx}
\usepackage{subcaption}
\usepackage{bm}

\newcolumntype{M}[1]{>{\centering\arraybackslash}m{#1}}
\newcolumntype{N}{@{}m{0pt}@{}}
\newcommand*\widefbox[1]{\fbox{\hspace{2em}#1\hspace{2em}}}

\numberwithin{equation}{section}
\usepackage{color}
\usepackage{float}
\include{MACROS_BK4}

\begin{document}

\title{A thermo-mechanically coupled finite deformation model for freezing-induced damage in soft materials}

\author[UNH]{Ali Saeedi}
\author[LSU]{Ram Devireddy}
\author[UNH]{Mrityunjay Kothari\corref{cor1}}
\ead{mrityunjay.kothari@unh.edu}
\cortext[cor1]{Corresponding author}
\address[UNH]{Department of Mechanical Engineering, University of New Hampshire, Durham, NH 03820, USA}
\address[LSU]{Department of Mechanical Engineering, Louisiana State University, Baton Rouge, LA 70803, USA}

\begin{abstract}
In the U.S., approximately 17 patients die each day awaiting an organ transplant, a crisis driven by the inability to store organs long-term via methods like cryopreservation. 
A primary failure mechanism is the severe thermo-mechanical damage tissues experience during freezing. 
A predictive understanding of this damage is hindered by the complex interplay between heat transfer, phase change, and large deformation mechanics.
Motivated by this fundamental problem, we present a fully coupled, thermo-mechanical phase-field framework for modeling damage evolution in fluid-saturated soft materials under cryogenic conditions. 
The theoretical framework integrates heat transfer with solid-liquid phase transition, finite deformation nonlinear elasticity, and progressive mechanical damage. 
The governing equations are solved using \texttt{FEniCS} finite element package.
The presentation will detail the theoretical framework and showcase representative simulations that capture the spatiotemporal evolution of temperature, freezing phase field, stress, and damage fields during representative freezing protocols. 
The developed framework serves as a powerful tool for understanding the fundamental mechanisms of freezing-induced injury and for designing improved cryopreservation strategies.
\end{abstract}

\maketitle

\section{Introduction}\label{introduction}%

Cryopreservation---the use of ultra-low temperatures to preserve biological tissues and organs---holds immense promise for extending organ transplantation windows, enabling tissue banking, and advancing regenerative medicine. 
However, the wide success achieved in cryopreservation of isolated cells \cite{RN75, RN66, RN46,RN60, RN57, RN62, RN65, RN63, RN41, RN28, RN99, rall1985ice, huang2018long, tessier2022partial, han2023vitrification, zhan2021cryopreservation} has not yet translated to the preservation of organs and tissues despite a dire need.
One of the \textit{grand challenges} to organ cryopreservation is that the ``larger'' size of the organs and tissues compared to single cells introduces a new and poorly understood form of damage caused by the excessive thermo-mechanical stresses, which leads to loss of biological function \cite{RN56, RN89, RN79, RN80, RN81, RN103, RN97, RN111, RN104, RN83, RN5}.\\

Most soft tissues contain $\sim 50$ to $\sim 70 \%$ water.
When tissues are frozen, the water undergoes phase change to ice progressively throughout the tissue; the frozen tissue then continues to cool down further until the desired temperatures are reached.
It is understood in the cryopreservation literature that different mechanisms of damage are at play in these two stages.
While in the first stage, dynamics of freezing/thawing, nucleation of ice crystals, and their interaction with cell membranes are the key contributors, in the second stage, bulk mechanical stresses due to phase-change induced volume change and differential thermal expansion/contraction due to temperature gradients are the major factors.

\begin{wrapfigure}{r}{12cm}
\centering
    \includegraphics[width=11.5cm]{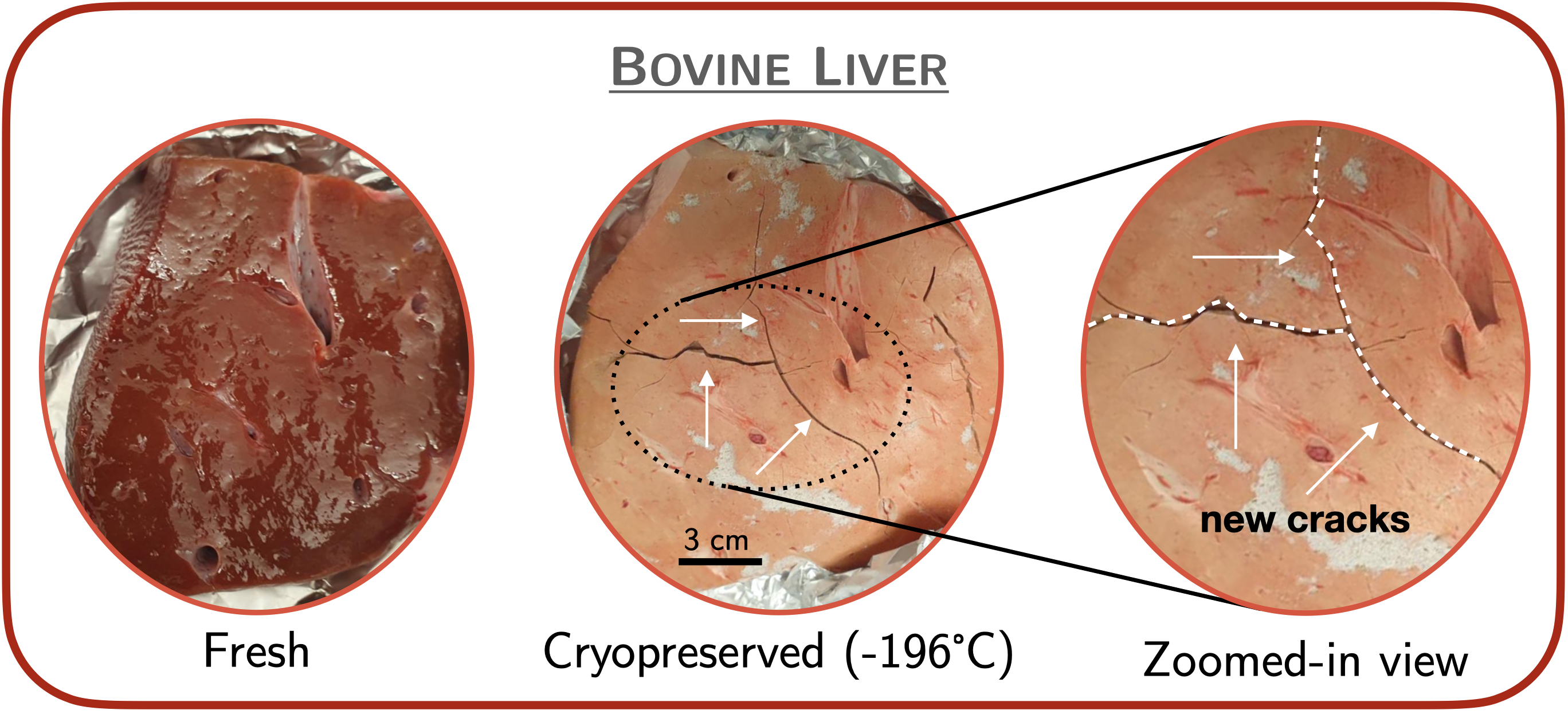}
 \caption{\footnotesize Thermal stresses induced cracking upon freezing a sample of bovine liver. Experimental image courtesy of Prof. Ram Devireddy's Bioengineering/Bioheat Transfer Laboratory at the Louisiana State University, Baton Rouge, LA.}\label{exp_cracks}
\end{wrapfigure}

These stresses, when severe, cause tissues to progressively lose structural integrity (``damage'') and ultimately develop visible cracks (see Figure \ref{exp_cracks}).\\

Given the immense promise of cryopreservation for human health, there has been an extensive body of work---both experimental and theoretical---which has significantly advanced the knowledge of this process.
We will confine our attention to the thermo-mechanical aspects of this process.
Early experiments on cryopreservation of animal tissues observed mechanical damage ranging from microscopic fractures \cite{adam1990effect, rabin1996experimental} to catastrophic gross fractures \cite{pegg1997fractures} during the entire cycle of freezing and thawing. 
Due to the non-uniform cooling of tissues (initiated from the boundaries and progressing to the interior), different regions of the tissues expand differently. 
Prior modeling efforts have indeed tested and confirmed this hypothesis and revealed important insights into the thermal stresses that develop during the freezing of tissues \cite{RN89, rabin1996analysis, RN80, RN81} and in heart geometries \cite{RN50}. 
In addition to reaffirming the experimental observations that thermal stresses are indeed high enough to push the tissues beyond their elastic limit, they also concluded that mechanical damage to the tissue is unavoidable no matter how slow the cooling is carried out \cite{RN103, RN97, RN111, RN104, RN83, RN5, RN34}. Studies in cylindrically symmetric settings showed that the volumetric expansion associated with the phase change of water ($\sim 9\%)$ is a significant contributor to the thermal stress (can elevate the maximum equivalent stress by $\sim$ 2-3x higher in renal tissues)\cite{he2005, shi1998thermal}. \\

Prior theoretical and computational efforts to model thermo-mechanical failures have largely focused on stress analyses and on avoiding fracture in simpler geometries, such as spheres and cylinders, to facilitate analytical tractability.
Furthermore, these studies are generally limited to simplified linear elastic or elastic-plastic models, which do not capture the non-linear elastic response that is typical of tissues.
Additionally, there is lack of theoretical and computational modeling that is able predict the evolution of damage due to thermo-mechanical stresses in tissues, which eventually leads to fracture. 
While fracture is certainly indicative of failure of the tissue and the cryopreservation protocol itself, damage is likely to be a more useful indicator of the extent of loss of biological function.
\\

Cryopreservation involves a confluence of thermal and mechanical processes that are intricately coupled, making it necessary to develop a fully coupled description of the processes, something that is currently missing from the literature.
Motivated by the importance of cryopreservation to human health and the existing mechanistic gaps in the literature, this works presents a fully-coupled, thermo-mechanically consistent framework for modeling freezing-induced damage in water-saturated soft materials, formulated within a finite deformation setting.
This effort will focus on developing a mechanistic picture of the second stage of the cryopreservation process, where bulk thermo-mechanical stresses play a crucial role.
\\

Freezing of tissues is a complex biophysical process.
Our approach to modeling is to develop the simplest model that captures all the relevant physical (thermo-mechanical) factors at play namely heat transfer, phase change, mechanical stresses and deformations, and the ensuing damage.
To this end, we integrate two phase-field models---one for fracture (or material damage) and one for the freezing process---into a unified continuum mechanics formulation. 
In contrast to prior cryopreservation models that relied on simplified geometries or linearized material behavior, our framework can accommodate complex geometries and incorporates a nonlinear elastic description of the soft tissue. 
The phase change from liquid water to ice is described by a continuously varying field that triggers volumetric expansion upon solidification, and this is coupled to the stress analysis so that thermal expansion/contraction and solidification-induced strain are naturally included. 
Similarly, the phase-field damage variable is coupled to the mechanical response, degrading the stiffness of the material as the material progressively loses its structural integrity under severe stresses.
By solving the strongly coupled heat transfer, phase transformation, and mechanical equilibrium equations, the model can capture the interplay between temperature evolution, ice formation, and mechanical damage in a way that is not possible with existing approaches. 
To our knowledge, this is the first theoretical framework that combines phase-field fracture and phase-field freezing in a finite-strain setting for soft biological materials. \\

The organization of the paper is as follows.
In \S2 we describe the thermo-mechanical theoretical framework, that integrates heat transfer, phase change, mechanical deformation/stresses, and damage.
The framework is then specialized in \S3 to a chosen set of free energies to derive the relevant governing equations.
The numerical implementation of the framework is validated through a selection of test problems in \S4.
The capabilities of the model are demonstrated by application of the specialized framework to four representative cases in \S5.
We conclude with a brief discussion of the results and future directions in \S6.

\section{Thermo-mechanical Theoretical Framework}
In this section we develop a coupled thermo-mechanical framework for modeling of evolution of damage in a freezing material over time.

\label{theory}

\subsection{Kinematic Descriptors}
We develop a theory that is based on the following kinematic descriptors: $\{\bfchi, \F, d, \nabla d, T, \nabla T,\phi, \nabla \phi\}$. 
Here, $\bfchi$ is the deformation field and $\F \coloneqq \nabla \bfchi$ is the deformation gradient; $0\leq d \leq 1$ is a damage phase field, which represents pristine material when $d=0$ and completely broken material when $d=1$;
 $T$ is the temperature; and, lastly, $0\leq \phi \leq 1$ is a freezing phase field which represents frozen material when $\phi =0$  and unfrozen when $\phi=1$.
The aim of the theory is to solve for $\{\bfchi, d, T, \phi\}$.

Since thermal expansion/contraction is a significant contributor to freezing stresses, we will adopt a  Kr\"oner-Lee multiplicative decomposition \citep{Lee1969} of the deformation gradient $\bfF$ into  elastic  and  phase change  parts $\bfF^e$ and $\bfF^{\rm ph}$ as

\begin{equation}\label{kroner-lee}
	\bfF = \bfF^e\bfF^{\rm ph}, \quad \text{where}	\quad J \equiv \det{\bfF}, \quad J^{\rm e}\equiv \det{\bfF^e} >0, \ \text{and} \ J^{\rm ph}\equiv\det{\bfF^{\rm ph}} >0.  
\end{equation}

In this work, we will consider $\F^{\rm ph} = (1+\epsilon^{T}(T, \phi)) \id$, with $\epsilon^{T}(T, \phi)$ being the transformation strain associated with thermal contraction/expansion and phase change.

\subsection{Macroscopic and Microscopic Force Balance} 
We employ the Principle of Virtual Power (PVP), following the approach of Germain \citep{germain1973method}, to derive the macroscopic and microscopic force balances involving our chosen kinematic variables and their associated power conjugates.

\noindent PVP requires that 
\begin{enumerate}
	\item \textit{The virtual power of all internal and external forces acting on a system, for any admissible virtual motion, be null;}
	\item \textit{The internal virtual power be an objective quantity.}
\end{enumerate}
Internal virtual power, $\mathcal{P}_i$ is:
\begin{equation}
	\mathcal{P}_i = \int_{\mathcal{B}} \left( \bfP:\dot{\F} +\omega \dot{d} + \bfzeta\cdot\nabla\dot{d} + \pi\dot{\phi} + \bfxi \cdot\nabla \dot{\phi}\right)dV_{\rm R},
\end{equation}
\begin{wrapfigure}{r}{8cm}
\centering
    \includegraphics[width=6cm]{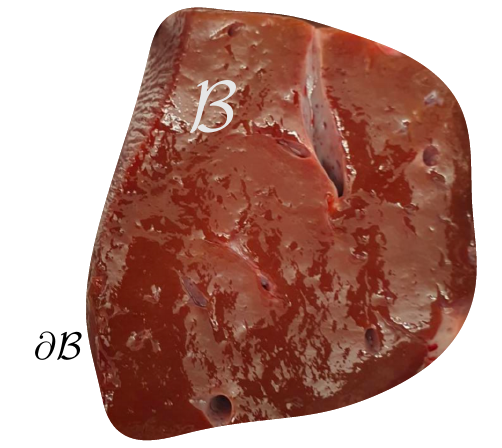}
 \caption{\small Schematic of the continuum body.}\label{wrap-fig2}
\end{wrapfigure}
where $\bfP$ (Tensorial Stress), $\omega$ (Scalar Microstress), $\bfzeta$ (Vector Microstress), $\pi$ (Scalar Microstress), $\bfxi$ (Vector Microstress) are defined as power conjugates to $\dot{\bfF}$ (distortion rate), $\dot{d}$ (rate of damage), $\nabla \dot{d}$ (gradient of rate of damage), $\dot{\phi}$ (rate of freezing), and $\nabla \dot{\phi}$ (gradient of rate of freezing) respectively.
The subscript $\rm R$ denotes the reference configuration and indicates that the above integral is expressed in reference configuration.
$\mathcal{B}$ denotes the continuum body and $\partial \mathcal{B}$ denotes its boundary (Fig. \ref{wrap-fig2}).\\

External virtual power, $\mathcal{P}_e$, is:
\begin{multline}
	\mathcal{P}_e = \int_\mathcal{B}\left(\bfb_{\rm R} \cdot \bfv + k\dot{d} + \bfl\cdot\nabla\dot{d} + \alpha \dot{\phi} + \bfbeta\cdot \nabla\dot{\phi} \right)dV_{\rm R} + \\
	\int_{\partial\mathcal{B}}\left( \bft_{\rm R}\cdot\bfv + h\dot{d} + \bfg \cdot \nabla\dot{d} + \gamma \dot{\phi} + \bfpi \cdot \nabla \dot{\phi}\right)dA_{\rm R},	
\end{multline}
where  $\bfb_{\rm R}$ is the body force per unit reference volume and $\bft_{\rm R}$ is the external mechanical traction that expend power on $\bfv$ (velocity); $k$ and $\alpha$ are microforces that expend power on $\dot{d}$ and $\dot{\phi}$ respectively;
$\bfl$ and $\bfbeta$ are microstresses that expend power on $\nabla \dot{d}$ and $\nabla \dot{\phi}$ respectively;
$h$ and $\gamma$ are microscopic tractions that expend power on $\dot{d}$ and $\dot{\phi}$ respectively;
$\bfg$ and $\bfpi$ are microscopic tractions that expend power on $\nabla \dot{d}$ and $\nabla \dot{\phi}$ respectively; all quantities are  defined in reference configuration.
\\

We require that $\mathcal{P}_{i}$ is objective under the transformation $\bfx^* = \bfy(t) + \bfQ(t)(\bfx - \bfo)$, where $\bfQ(t) \in \mathrm{SO}(3)$ is an orthogonal rotation matrix, which gives the following transformation rules:
\begin{subequations}
\begin{align}
		\bfP^{*} =& \ \bfQ \bfP, \quad \bfP \F^{\top} = \bfP^{\top}\F \\
	\omega^* =& \ \omega, \quad 	\pi^* =\ \pi,\\
	\bfzeta^* =& \ \bfQ\bfzeta, \quad 	\bfxi^* =\ \bfQ\bfxi
\end{align}
\end{subequations}
Finally, from $\mathcal{P}_{i} = \mathcal{P}_{e}$, we obtain the following macroscopic and microscopic balances\footnote{In this calculation, it is implicitly assumed that there are no external volume fields that expend power on the $\dot{d}$,  $\nabla\dot{d}$, $\dot{\phi}$, and $\nabla\dot{\phi}$, thus nullifying $k$, $\bfl$, $\alpha$, and $\bfbeta$.}
\begin{subequations}
\begin{align}
  \nabla\cdot \bfP +  \bfb = &0,\ \X \in \mathcal{B} & \text{Balance of Lin. Mom. \footnotemark} \label{bolm}\\
     \bfP \n_{\rm R} - \bft_{\rm R}=&0, \ \X \in \partial\mathcal{B} & \text{Cauchy's Law} \label{cauchy-law}\\
   \nabla\cdot \bfzeta -\omega =&0, \ \X \in \mathcal{B}  & \text{Microforce Balance (for $d$)} \label{damage-evol}\\
    \bfzeta \cdot \n_{\rm R} -h = &0, \ \X \in \partial\mathcal{B} & \text{Microscopic Traction (for $d$)} \\
   \nabla\cdot \bfxi -\pi =&0, \ \X \in \mathcal{B}  & \text{Microforce Balance (for $\phi$)}\label{freezing-evol}\\
    \bfxi \cdot \n_{\rm R} -\gamma= &0, \ \X \in \partial\mathcal{B}  &\text{Microscopic Traction (for $\phi$)} \\
        \bfg = &0, \ \X \in \partial\mathcal{B}  &\text{Microscopic Traction (for $d$)} \\
            \bfpi = &0, \ \X \in \partial\mathcal{B}  &\text{Microscopic Traction (for $\phi$)}
\end{align}
\end{subequations}
\footnotetext{Readers familiar with this process will recognize that $\bfP$ is the First Piola Kirchhoff Stress Tensor.}

With these equations, the final form of the virtual power balance becomes:
\begin{empheq}[box=\widefbox]{align}
	\int_{\mathcal{B}} \left( \bfP:\dot{\F} +\omega \dot{d} + \bfzeta\cdot\nabla\dot{d} + \pi\dot{\phi} + \bfxi \cdot\nabla \dot{\phi}\right)dV_{\rm R} =  \int_\mathcal{B} \bfb_{\rm R} \cdot \bfv \ dV_{\rm R} + 
	\int_{\partial\mathcal{B}}\left( \bft_{\rm R}\cdot\bfv + h\dot{d}  + \gamma \dot{\phi}\right)dA_{\rm R}.\label{pvp_balance}
\end{empheq}

\subsection{First and Second Laws of Thermodynamics}
For thermodynamic consistency, the theory must satisfy the \textit{First} and \textit{Second Laws of Thermodynamics}.
Following the internal virtual power form in \eqref{pvp_balance}, the First Law of Thermodynamics gives the following balance of energy in $\mathcal{B}$, 
\begin{equation}\label{first_law}
	\dot{e} = \bfP \colon \dot{\bfF} + \omega \dot{d} +\bfzeta\cdot \nabla \dot{d} + \pi\dot{\phi} + \bfxi\cdot\nabla\dot{\phi}+ r - \nabla \cdot \bfq_{\rm R},
\end{equation}
where $e$ is the internal energy per unit reference volume, $r$ is some volumetric internal heating source per unit reference volume, and $\bfq_{\rm R}$ is the heat-flux vector expressed in the referential form.
The remaining terms correspond to the power expended on the system by various stresses, microforces, and microstresses.

\paragraph{Aside on Stress Power.} We also note here that in view of equation \eqref{kroner-lee}, $\dot{\F}, \dot{\F}^{\rm e},$ and $\dot{\F}^{\rm ph}$ are related.
Thus, the term $\bfP:\dot{\F}$ can be split into the following terms \citep{chester2011thermo},
\begin{equation}
	\bfP:\dot{\F} = J\bfsigma \F^{-\rm e\top}\colon \dot{\F}^{e} + J\F^{\rm e\top}\bfsigma\F^{-\rm e\top}\colon\dot{\F}^{\rm ph}{\F}^{\rm ph-1}
\end{equation}
where $\bfsigma = J^{-1}\bfP \F^{\top}$ is the Cauchy Stress tensor.
Since the strain energy functions are typically expressed as functions of the elastic right Cauchy-Green tensor $\C^{e}$, the above equation can be recast following the approach in \citep{chester2011thermo} as,
\begin{equation}\label{stress_power_alternative}
	\bfP:\dot{\F} = \frac{1}{2}\bfsigma^{\rm e}\colon \dot{\C}^{e} -\bar{p} \dot{J}^{\rm ph}
\end{equation}
where 
\begin{align}
	\bfsigma^{\rm e} = & \F^{\rm e-1} J \bfsigma \F^{\rm e-\top}=\F^{\rm e-1} \bfP \F^{\rm ph\top}, \\
	\C^{\rm e} =& \F^{\rm e \top} \F^{\rm e}, \quad \text{and}\\
	\bar{p} =&-\frac{1}{3}J^{\rm e} {\rm tr}(\bfP \F^{\top}/J)
\end{align} \\

By introducing the Helmholtz free energy $\psi = e-Ts$, where $s$ is the entropy per unit reference volume, we can rearrange the internal energy equation \eqref{first_law}, using \eqref{stress_power_alternative}, as
\begin{equation}\label{entropy_evolution}
T\dot{s} = \frac{1}{2}\bfsigma^{\rm e}\colon \dot{\C}^{e} + \omega \dot{d} +\bfzeta\cdot \nabla \dot{d} + \left(\pi-\bar{p}\pards{J^{\rm ph}}{\phi}\right)\dot{\phi} + \bfxi\cdot\nabla\dot{\phi}+ r - \nabla \cdot \bfq_{\rm R} - \dot{\psi} -\dot{T}\left(s + \bar{p}\pards{J^{\rm ph}}{T}\right)
\end{equation}
where we have used $\small \dot{J}^{\rm ph} = \pards{J^{\rm ph}}{\phi}\dot{\phi}+\pards{J^{\rm ph}}{T}\dot{T}$.

The Second Law of Thermodynamics states that the total entropy change of the system and surroundings must be non-negative.
This translates into the following requirement, also known as the \textit{Dissipation Inequality}:
\begin{equation}\label{disspiation_inequality}
\mathcal{D} \equiv \frac{1}{2}\bfsigma^{\rm e}\colon \dot{\C}^{e} + \omega \dot{d} +\bfzeta\cdot \nabla \dot{d} + \left(\pi-\bar{p}\pards{J^{\rm ph}}{\phi}\right)\dot{\phi}+ \bfxi\cdot\nabla\dot{\phi} - \dot{\psi} -\dot{T}\left(s + \bar{p}\pards{J^{\rm ph}}{T}\right) - \frac{\bfq_{\rm R}\cdot \nabla T}{T} \geq 0.
\end{equation}

Assuming a general form for the Helmholtz free energy as $\psi \equiv \psi(\bfC^{\rm e},d, \nabla d, T, \nabla T,\phi, \nabla \phi)$, the $\dot{\psi}$ term can be expanded as follows,
\begin{equation}\label{psidot}
	\dot{\psi} = \pards{\psi}{\bfC^{e}}\colon \dot{\bfC}_e + \pards{\psi}{d}\ \dot{d} +\pards{\psi}{\nabla d}\cdot \nabla \dot{d} +\pards{\psi}{T}\ \dot{T}+\pards{\psi}{\phi}\ \dot{\phi} + \pards{\psi}{\nabla \phi} \cdot\nabla\dot{\phi} , 
\end{equation}
which yields the following concise form for the dissipation inequality,
\begin{multline}
\label{disspiation_inequality2}
\mathcal{D} \equiv \left(\frac{1}{2}\bfsigma^{\rm e} - \pards{\psi}{\bfC^{e}}\right) \colon \dot{\bfC}_e + \left(\omega -\pards{\psi}{d}\right) \dot{d} +\left(\bfzeta-\pards{\psi}{\nabla d}\right)\cdot \nabla \dot{d} +\left(\pi-\bar{p}\pards{J^{\rm ph}}{\phi} - \pards{\psi}{\phi} \right)\dot{\phi}  \\ +\left(\bfxi - \pards{\psi}{\nabla \phi}\right) \cdot\nabla\dot{\phi} - \left(s + \bar{p}\pards{J^{\rm ph}}{T}+\pards{\psi}{T} \right)\dot{T} - \frac{\bfq_{\rm R}\cdot \nabla T}{T} \geq 0.	
\end{multline}

\subsection{Constitutive Relations}

Through the application of \textit{Coleman-Noll} procedure \citep{noll1974thermodynamics} on the chosen form of free energy, the dissipation inequality yields the following equations:
\begin{subequations}
\begin{align}
  	 \bfsigma^{\rm e} =& 2 \pards{\psi}{\mathbf{C}^{\rm e}} & \text{Stress}\label{constitutive_eqns-1} \\
   s =& -\bar{p}\pards{J^{\rm ph}}{T}-\pards{\psi}{T}  & \text{Entropy Density} \label{constitutive_eqns-2}
\\
    \omega =& \pards{\psi}{d}  &\text{Scalar Microstress} \\
    \pi = & \underbrace{\bar{p}\pards{J^{\rm ph}}{\phi}+\pards{\psi}{\phi}}_{\pi_{\rm energetic}} + \underbrace{\frac{1}{M}\dot{\phi}}_{\pi_{\rm dissipative}}   &\text{Scalar Microstress}\label{constitutive_eqns-4} \\
    \bm{\zeta} =&  \pards{\psi}{\nabla d}  &\text{Vector Microstress}\\
    \bm{\xi} = & \pards{\psi}{\nabla \phi}  &\text{Vector Microstress}\\
    \bfq_{\rm R} =& -K_{\rm R}({\bfC}^{\rm e}, d,\phi, T)\nabla T &\text{Heat Conduction} \label{constitutive_eqns-6}
  \end{align}
\end{subequations}
In general, any choice that is consistent with \eqref{disspiation_inequality2} is a valid choice.
In the above application of the Coleman-Noll procedure, we make a specific constitutive choice that $\pi$ has an energetic and a dissipative part.
This will be complemented with an evolution equation for $\phi$ in the next subsection.
$M > 0$ is a mobility constant, and $K_{\rm R}({\bfC}^{\rm e},d,\phi, T)>0$ is the thermal conductivity.

\subsection{Damage-informed Temperature and Freezing Evolution} 
\paragraph{Temperature Evolution.} Starting from the entropy evolution equation \eqref{entropy_evolution} and using equations \eqref{psidot} and \eqref{constitutive_eqns-1}-\eqref{constitutive_eqns-6}, we get the following simplified equation for entropy evolution
\begin{equation}\label{entropy_evolution2}
	T\dot{s} = \pi_{\rm diss.} \dot{\phi} +  r - \nabla \cdot \bfq_{\rm R}.
\end{equation}
Recalling from equation \eqref{constitutive_eqns-2} that $s =-\bar{p}\pards{J^{\rm ph}}{T}-\pards{\psi}{T}$, the entropy evolution equation can be recast in the following form for temperature evolution
\begin{equation}
\underbrace{T\pards{s}{T}}_\text{$C_{v}$, Volumetric Heat Capacity}\dot{T}= r-\nabla \cdot \bfq_{\rm R} \ + \ \underbrace{\pi_{\rm diss.}\dot{\phi} - T\left(\pards{s}{\bfC^{\rm e}}:\dot{\bfC}^{\rm e} + \pards{s}{d}:\dot{d} + \pards{s}{\nabla d}:\nabla\dot{d}+\pards{s}{\phi}:\dot{\phi} + \pards{s}{\nabla \phi}:\nabla\dot{\phi}\right)}_\mathcal{L} .\label{temp_evolution}
\end{equation}

\paragraph{Freezing Phase Field Evolution.} From the microscopic balance law \eqref{freezing-evol}  and scalar microstress equation \eqref{constitutive_eqns-4}, we obtain
\begin{equation}
	\nabla \cdot \pards{\psi}{\nabla\phi} = \pi_{\rm energetic} + \pi_{\rm diss.},\\
\end{equation}
which gives the following form for $\pi_{\rm diss.}$
\begin{equation}
	\pi_{\rm diss.}  = \nabla \cdot \pards{\psi}{\nabla\phi}  -\bar{p}\pards{J^{\rm ph}}{\phi}-\pards{\psi}{\phi}
\end{equation}
Thus, the evolution equation for the freezing phase field becomes,
\begin{equation}
	\dot{\phi} = -M\left(\bar{p}\pards{J^{\rm ph}}{\phi}+\pards{\psi}{\phi}  - \nabla \cdot \pards{\psi}{\nabla\phi}\right).
\end{equation}

\subsection{Summary of Equations}
The relevant governing equations for a rate-independent model are\footnotemark\footnotetext{Here, it is assumed that mechanical fields equilibriate much faster than thermal fields. $\mathcal{H}$ is a history variable introduced to ensure that the crack \textit{does not} heal.}:
\begin{empheq}[box=\widefbox]{align}
	\nabla \cdot \bfP  + \bfb &= 0 \quad  &\text{Mechanical Equilibrium}\\
	\nabla\cdot  \pards{\psi}{\nabla d} -\pards{\psi}{d} &=0 \quad  &\text{Damage Evolution} \\
	 c \dot{T} -\mathcal{L} &= r -\nabla \cdot \bfq_{\rm R} \quad  &\text{Heat Conduction}\\
	\dot{\phi} + M\left(  \bar{p}\pards{J^{\rm ph}}{\phi}+\pards{\psi}{\phi}    - \nabla \cdot \pards{\psi}{\nabla\phi}\right) &=0\quad  &\text{Freezing Phase Field Evolution}
\end{empheq}
supplemented with appropriate initial and boundary conditions for $\bfchi, d$,  $T$, and $\phi$.
With proper choice of the constitutive equations, the model applies to rate-dependent damage as well.

\section{Specialization of Framework}
As an illustrative example, we will specialize the general model developed in \S 2 to specific choices of free energies and derive the relevant governing equations.
We choose the following form for the free energies for this exercise:
\begin{equation}\label{free_energy}
\psi(\C^{\rm e}, d, \nabla d, T, \phi, \nabla\phi) = \psi_{\rm el}(\C^{\rm e}, d) + \psi_{\rm crack}(d, \nabla d, \phi) + \psi_{\rm thermal}(T, \phi, \nabla \phi),
\end{equation}
where the first term is the elastic strain energy stored in the material, the second term is the diffused fracture (or damage) energy (equivalent to $\int G_c dA$ in Griffith's theory \citep{griffith1921vi}), and the last term is the thermal energy.
In the following, we will specialize each of these energies to a specific form.
\subsection{Nonlinear elasticity: slightly compressible neo-Hookean model}
The elastic free energy for a neo-Hookean model is chosen to be
\begin{equation}
\psi_{\rm el}(\bfC^{e}, d) = g_d(d)\left[\frac{\mu}{2}\left(\rm \bar{I}_1 -3 \right) + \frac{K}{2}(\ln{J^{\rm e}})^2\right]
\end{equation}
where $g_d(d) = (1-d)^2$ is a degradation function that \textit{degrades} the elastic energy density (more discussion in the next subsection), $\bar{I}_1 = \rm tr{\bar{\bfC}^{e}}$ and $\bar{\bfC}^{e} = J^{\rm e-2/3}\bfC^{e}$.
From the constitutive relation \eqref{constitutive_eqns-1}, we obtain
\begin{equation}
	\bfsigma^{\rm e} = g_d(d)\left[\mu  J^{\rm e-2/3}\left( \id -\frac{1}{3}\bfC^{\rm e-\top}\rm tr{\Ce}\right) + K \ln{J^{\rm e}}\bfC^{\rm e-\top}\right].
\end{equation}
Cauchy and Piola-Kirchoff stresses are then obtained as,
\begin{align}
	J\bfsigma =& g_d(d) \left[\mu \dev(\bar{\bfB}^{e}) + K \ln J^{\rm e}\id\right], \quad \text{and}\\
	\bfP \equiv &g_d(d) \left[\mu \dev(\bar{\bfB}^{e})  -J^{\rm e}p\id\right]\bfF^{\rm -\top} =  g_d(d) \left[\mu \dev(\bar{\bfB}^{e}) + K \ln J^{\rm e}\id\right]\bfF^{\rm -\top}.
\end{align}
where $p =-K \ln J^{\rm e}/J^{\rm e} $.
Furthermore, we get
\begin{equation}
	\bar{p} = -g_d(d)\frac{J^{\rm e}}{J}K \ln J^{\rm e}.
\end{equation}
We will consider $\F^{\rm ph} = (1+\epsilon^{T}(\phi, T)) \id$ for this specialization; the functional form is given in \S3.3.

\subsection{Damage}
A number of damage models have been proposed in the literature that \textit{smear} the crack over a finite width and thus regularize the sharp interface crack problem \cite{bourdin2000numerical, bourdin2008variational, lopez2025classical, kumar2020revisiting, kumar2020phase, miehe2010thermodynamically, raina2016phase, konale2025modeling, konale2025physics, talamini2018progressive, mao2018theory, alkhoury2025finite}\footnote{We do not attempt to present an exhaustive literature review of damage modeling in this work.}.
There are recent works in the literature that have proposed theories that unify nucleation and crack propagation in a single framework by means of a strength surface.
On the other hand, classical models have proposed choosing the length scale such that the tensile strength of the material matches the localization stress from the models.
The focus of the current work is not to propose a new damage model or comment on their relative strengths, but on the overall framework that integrates the various physical processes together.
To this end, the framework presented in \S2 can be adapted to any suitable choice of damage model appropriate for the specific material and problem at hand.
Here, for purposes of illustration of the framework, we will choose the \texttt{AT2} model where the free energy associated with cracking is,
\begin{equation}
\psi_{\rm crack} = \int_{\mathcal{B}}G_c \gamma(d, \nabla d)dV, 
\end{equation}
where 
\begin{equation}
	\gamma(d, \nabla d) = \frac{w(d)}{2l} + \frac{l}{2}|\nabla d|^2
\end{equation}
is the \textit{crack surface density} function, $w(d) = d^2$ is a \textit{geometric function} that controls local fracture energy, and $G_c$ is the critical energy release rate and is assumed to be a material property.
In the limit of the regularization length $l \rightarrow 0$, the minimization problem converges to the classical sharp-crack problem.
In this specialization, we have chosen a simple degradation scheme that degrades both the isochoric and volumetric parts of the energy.
A more sophisticated split can be easily achieved in this framework.
For the case of tissues in particular, biological damage can happen even under compression which may lead to mechanical damage (loss in ability to sustain loads), thus an appropriate split of the energy is an open research question in itself.

\subsection{Phase Change}
\begin{figure}[h!]
  \centering
  \begin{subcaptionbox}{\label{fig:sub1}}[0.31\textwidth]
    {\includegraphics[scale=0.23]{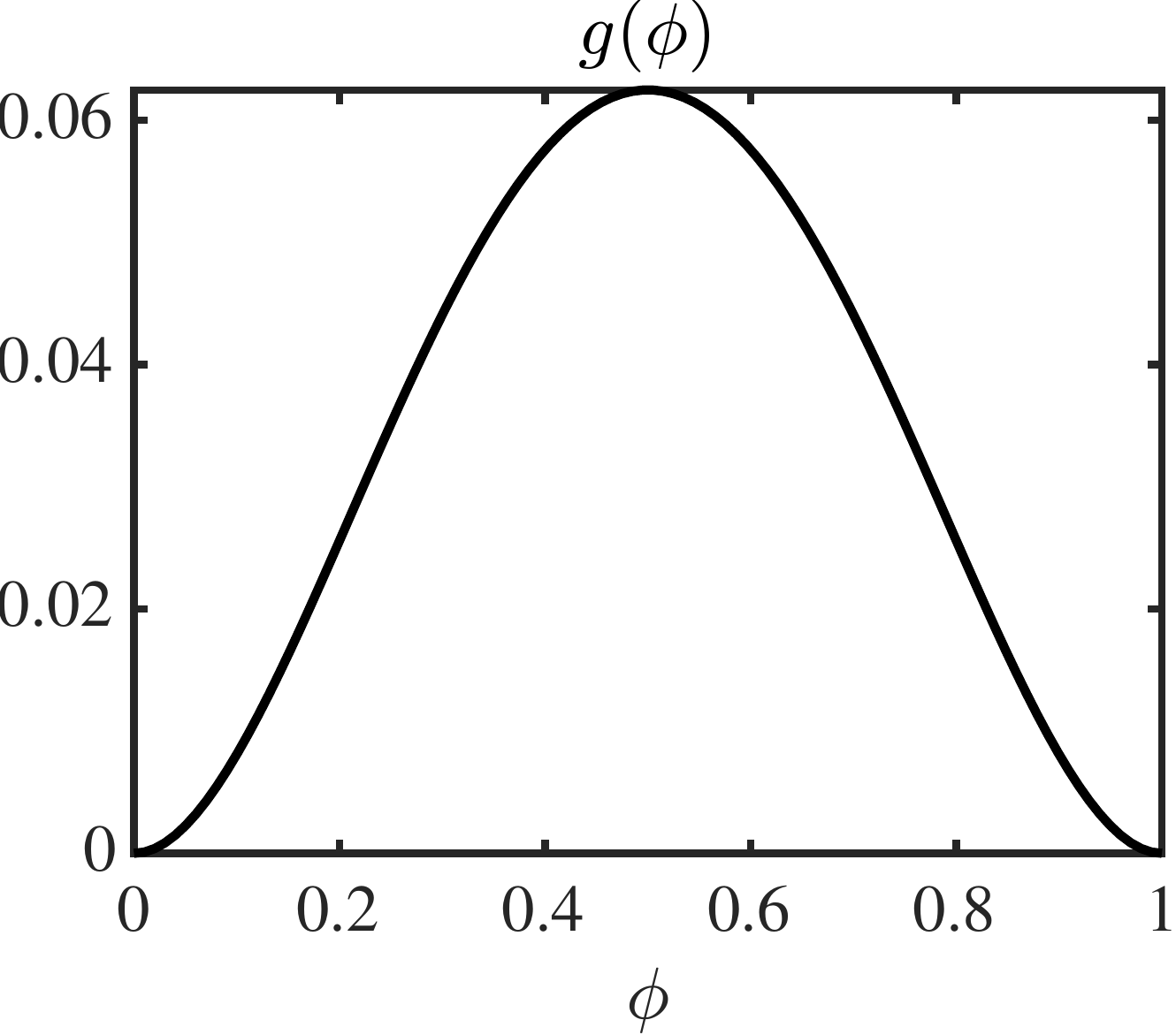}}
  \end{subcaptionbox}
  \hfill
  \begin{subcaptionbox}{\label{fig:sub2}}[0.32\textwidth]
    {\includegraphics[scale=0.23]{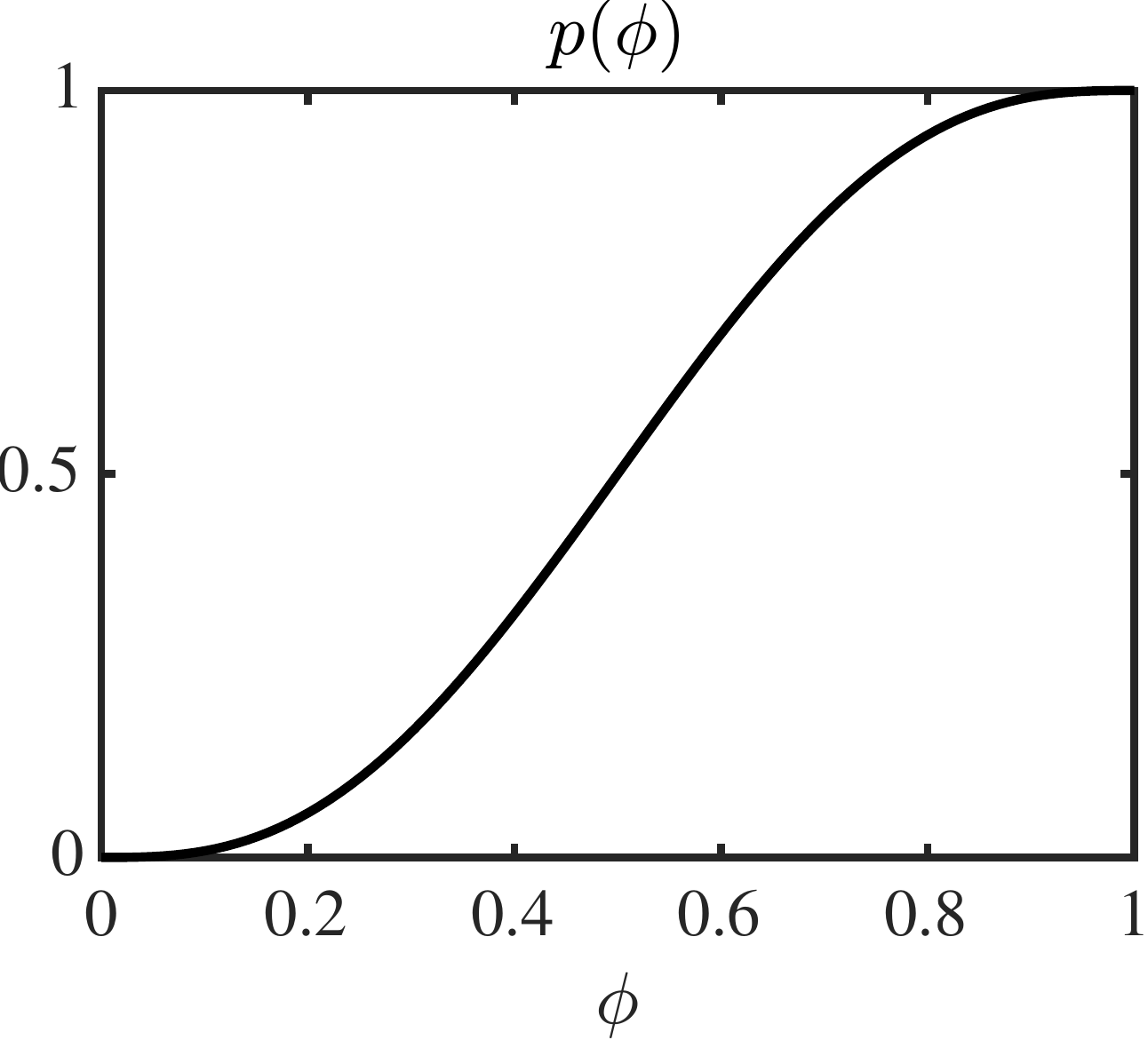}}
  \end{subcaptionbox}
  \hfill
  \begin{subcaptionbox}{\label{fig:sub3}}[0.32\textwidth]
    {\includegraphics[scale=0.23]{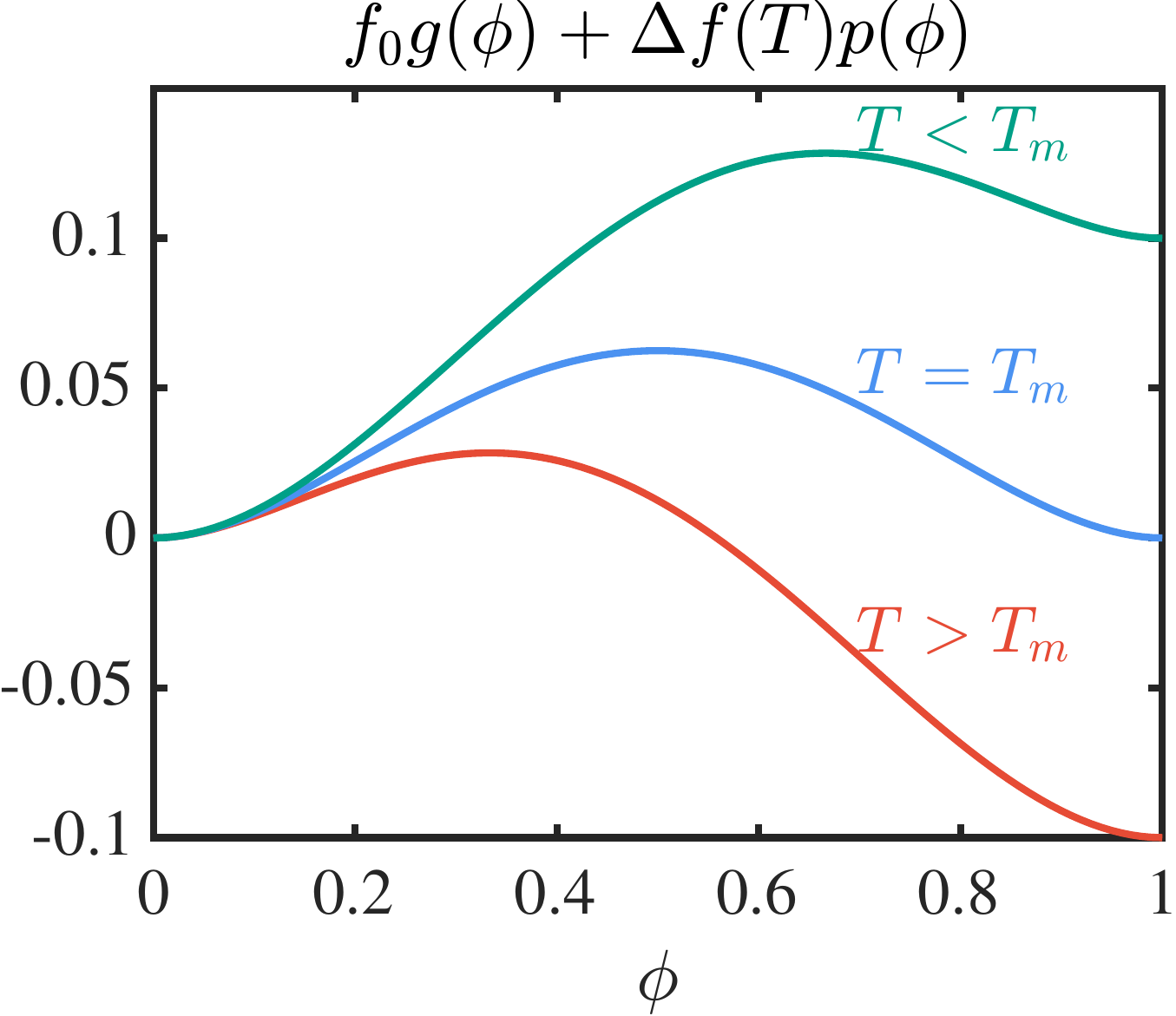}}
  \end{subcaptionbox}
  \caption{(a) Double-well potential $g(\phi)$; (b) smooth interpolation function $p(\phi)$; (c) schematic of thermal free energy at different temperatures.}
  \label{doublewell}
\end{figure}

The solidification phase field $\phi \in [0,1]$ is defined such that $\phi=0$ represents the frozen phase while $\phi=1$ represents the unfrozen phase.
We choose the thermal free energy, $\psi_{\rm thermal}$ as comprising a symmetric double-well potential function that presents an energy barrier to freezing and a temperature-dependent latent heat term that introduces asymmetry in the free energy.
\begin{equation}
\psi_{\rm thermal} = f_{\rm 0} g(\phi) + \Delta f(T) p(\phi) + \frac{\beta}{2} |\nabla \phi|^2
\end{equation}
where $\Delta f(T) = L(T_m-T)/T_m$, $g(\phi) = \phi^2(1-\phi)^2$, $p(\phi) = \phi^3(6\phi^2-15\phi +10)$, and $T_m$ is the melting temperature.
For these choices, $p'(\phi) = 30g(\phi)$.
Finally, for transformation strain for phase change, we choose $\epsilon^{T}(\phi, T) = \epsilon_0(1-p(\phi)) + \alpha(\phi)(T-T_m)$, where $\epsilon_0$ is the transformation strain associated with phase change from water to ice, and $\alpha(\phi) = \phi \alpha_{\rm water} + (1-\phi)\alpha_{\rm ice}$ is the effective coefficient of thermal expansion.
Figure \ref{doublewell} shows a sketch of the various functional forms.

\subsection{Evolution Equations}
Entropy for the chosen functional forms of free energies becomes
\begin{equation}
	s = -\bar{p}\pards{J^{\rm ph}}{T}-\pards{\psi}{T} = 3 J^{\rm ph -1/3}Kg_d(d) \ln J^e \alpha(\phi)+\frac{L}{T_m}p(\phi)
\end{equation}
which yields the following evolution equation for the temperature field \eqref{temp_evolution}
\begin{multline}\label{temp_evol}
	C_v\dot{T} = K_{\rm R}\nabla^2 T + \frac{1}{M}\dot{\phi}^2 - \underbrace{\frac{3}{2}TKg_d(d)J^{\rm ph -1/3} \alpha(\phi) \bfC^{e-\top}}_{T\pards{s}{\Ce}}\colon \Cedot -\\
	 \underbrace{T\left[ -3J^{\rm ph -2/3}\alpha(\phi)\pards{\epsilon^{T}}{\phi}Kg_d(d) \ln J^e +3J^{\rm ph -1/3}\alpha'(\phi)Kg_d(d) \ln J^e + \frac{L}{T_m}p'(\phi)\right]}_{T\pards{s}{\phi}}\dot{\phi} - \\
	 \underbrace{T\left[3J^{\rm ph -1/3}Kg_d'(d) \ln J^e \alpha(\phi) \right]}_{T\pards{s}{d}}\dot{d}
\end{multline}
where $C_v$ is the volumetric heat capacity (J/m$^3$), and $L$ is the latent heat of fusion (J/m$^3$)\footnote{For the specific choices made in the specialization, a number of terms that involve partial derivatives with $T$ vanish from the latent heating in \eqref{temp_evolution}.}, \footnotemark.

\footnotetext{$\frac{1}{M}\dot{\phi}^2$ is a valid term that shows in this formulation, as also reported in \cite{fried1993continuum}. However, we will drop this term under the assumption that kinetics are sufficiently slow, which will make the contribution of this term of lower order than other terms.}
The freezing phase field $\phi$ evolves according to the following equation,
\begin{equation}
	\dot{\phi} = -M\left(\bar{p}\pards{J^{\rm ph}}{\phi} + f_{\rm 0}g'(\phi) + L\frac{T_m-T}{T_m}p'(\phi) -\beta \nabla^2\phi\right)
\end{equation} \label{phi_evol}

Using 
\begin{equation}
	J^{\rm ph} =  (1+\epsilon^T(\phi, T))^3
\end{equation} and 
\begin{equation}
	\epsilon^{T}(\phi, T) = \epsilon_0(1-p(\phi)) + \alpha(\phi)(T-T_m),
\end{equation}
we obtain that 
\begin{equation}
	\pards{J^{\rm ph}}{\phi} =  3J^{\rm ph 2/3}(-p'(\phi)\epsilon_0 + \alpha'(\phi)(T-T_m))
\end{equation}

The evolution equation for the freezing phase field then takes the following form,
\begin{equation}
	\dot{\phi} = -M\left(3p'(\phi)\epsilon_0J^{\rm ph -1/3}Kg_d(d) \ln J^{\rm e} -3\alpha'(\phi)(T-T_m)J^{\rm ph -1/3}K g_d(d)\ln J^{\rm e} + f_{\rm 0}g'(\phi) + L\frac{T_m-T}{T_m}p'(\phi) -\beta \nabla^2\phi\right).
\end{equation} \label{phi_evol}

\noindent The mechanical behavior of the system is assumed to take place at a much faster timescale, and thus it is always assumed to be in equilibrium.
These equations are
\begin{align}
	\nabla \cdot \bfP &=0 \\
	G_cl\nabla^2 d +2(1-d)\mathcal{H}-\frac{G_c}{l} d &=0 \quad  &\text{Damage Evolution} \\
	\mathcal{H} &= \max(\mathcal{H}_{\rm old}, \psi_{\rm el}) \quad  &\text{History Variable}\label{history_var}
\end{align}
\section{Benchmarking of the Model and Numerical Implementation} \label{benchmarking}
In this section we will test the various modules of our model and numerical implementation by applying it to simplified problems whose solutions can be obtained analytically.

\subsection{Numerical Implementation}

The coupled thermomechanical phase-field damage model is implemented within the\texttt{FEniCS}finite element framework \cite{AlnaesEtal2015, LoggEtal2012, LoggWells2010, LoggEtal_10_2012, KirbyLogg2006, LoggEtal_11_2012, OlgaardWells2010, Kirby2004, kirby2010}.

\paragraph{Discretization.}
The computational domains considered in \S4 \& 5 are discretized either using meshes generated natively within\texttt{FEniCS}or using an external mesh generator ``Gmsh'' \cite{gmsh} and imported into the solver via the XDMF format. 
A mixed finite element method was employed for spatial discretization. 
The displacement field $\mathbf{u}$ was approximated using continuous second-order (P2) Lagrange elements, while the pressure $p$, temperature $T$, and phase-field $c$ were approximated with continuous first-order (P1) elements to form a stable Taylor-Hood-like element for the $(\mathbf{u}, p)$ pair. 
The damage field $d$ and damage rate $\dot{d}$ were solved in a separate P1 function space. Time integration was performed using a first-order, implicit backward-Euler scheme.

\paragraph{Solution Strategy.}
A staggered (operator-split) solution strategy \cite{miehe2010phase} was employed at each time step to handle the multi-physics coupling. The iterative procedure within a time step proceeds as follows:
\begin{enumerate}
    \item First, the monolithic coupled system for the primary fields $(\mathbf{u}, p, T, c)$ was solved. This step incorporates the thermal source term from damage, utilizing the damage rate $\dot{d}$ calculated from the previous staggered iteration.
    \item Next, the elastic energy density $\psi_{\rm el}$ was computed to serve as the damage driving force $\mathcal{H}$ in accordance with \eqref{history_var}.
    \item Using this updated history field, the separate non-linear damage evolution problem for $d$ was solved.
    \item Finally, the damage rate field $\dot{d}$ was updated using the new damage solution.
\end{enumerate}
This process was iterated until the $L^2$-norm of the change in the damage field fell below a relative tolerance of $10^{-4}$. All non-linear systems were solved using a Newton-Raphson method, and resulting linear systems were resolved via the MUMPS parallel direct sparse solver.

\subsection{Finite Deformation: Uniaxial loading with prescribed displacement}
\begin{figure}[h!]
  \centering
  \begin{subcaptionbox}{\label{fig:sub1_uniaxial}}[0.31\textwidth]
    {\includegraphics[scale=0.65]{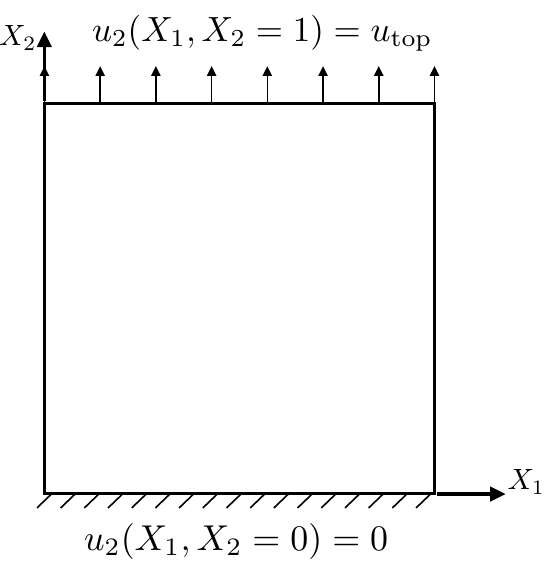}}
  \end{subcaptionbox}
\hspace{50pt}
  \begin{subcaptionbox}{\label{fig:sub2_uniaxial}}[0.32\textwidth]
    {\includegraphics[scale=0.3]{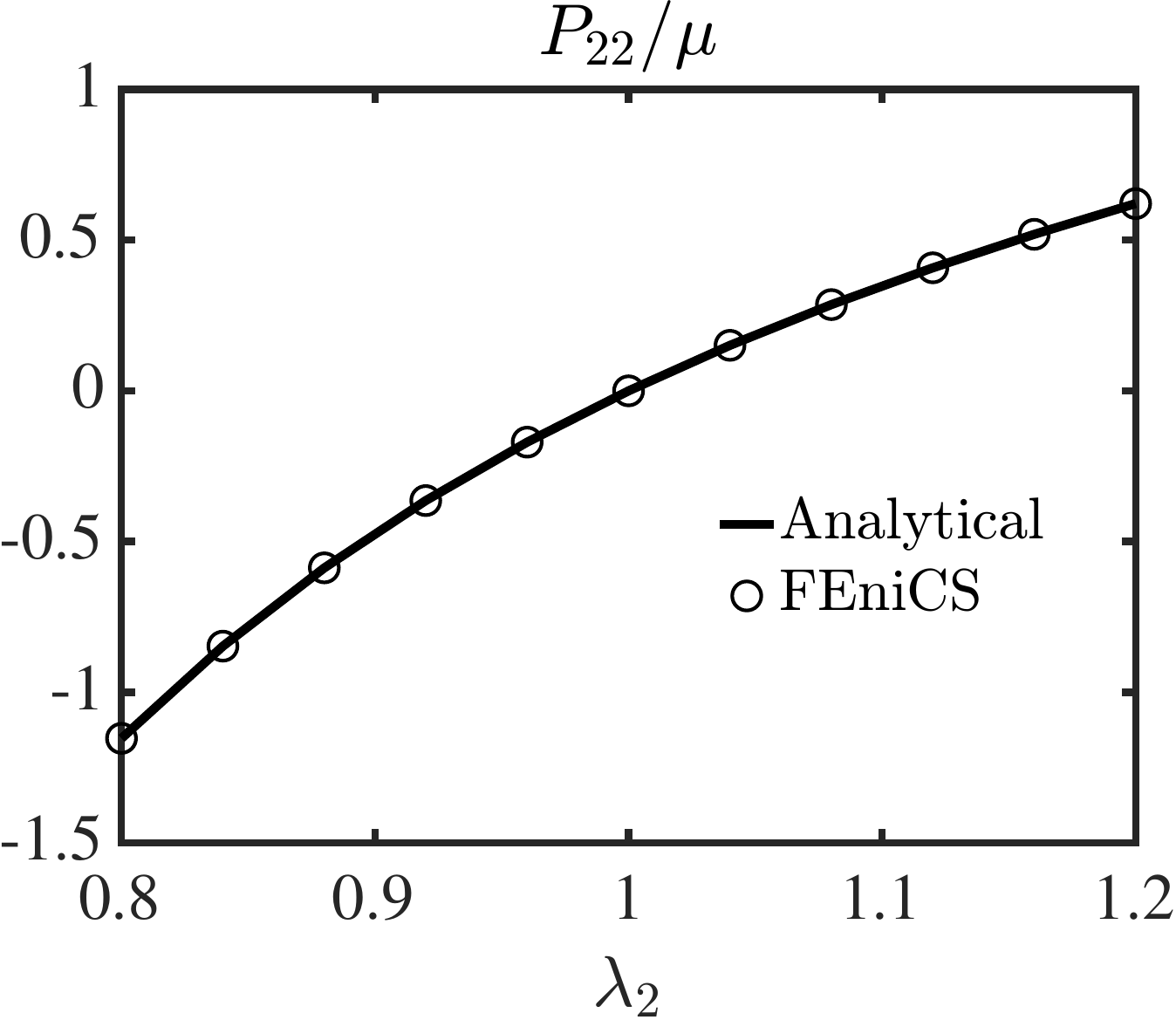}}
  \end{subcaptionbox}
  \hfill
  \caption{(a) Schematic of the setup to study uniaxial loading of a nearly incompressible, plane strain neo-Hookean sample. The domain was discretized natively in\texttt{FEniCS}using 40,000 structured quadrilateral elements. (b) First Piola Kirchhoff stress, $P_{22}$, obtained from simulation matches the analytical results.}
  \label{uniaxial_schematic}
\end{figure}

We consider a 2D, unit square with $u_2(X_1, X_2=0) =0$ and $u_2(X_1, X_2=1)=$\texttt{u$_{\texttt{top}}$}, as shown in the figure \ref{uniaxial_schematic}.
No other displacements are constrained (other than pinning the bottom left corner to prevent translation) and the left and right boundaries are traction-free.
The analytical solution for this problem for a nearly incompressible neo-Hookean material under plane strain conditions is given as,
\begin{align}
	P_{11} = & \frac{\mu}{(\lambda_1\lambda_2)^{2/3}} \left[\frac{2\lambda_1}{3}-\frac{\lambda_2^2}{3\lambda_1}  -\frac{1}{3\lambda_1}\right] + K\frac{\ln(\lambda_1\lambda_2)}{\lambda_1} = 0, \label{P11}\\
	P_{22} =& \frac{\mu}{(\lambda_1\lambda_2)^{2/3}} \left[-\frac{\lambda_1^2}{3\lambda_2} +\frac{2\lambda_2}{3} -\frac{1}{3\lambda_2}\right] + K\frac{\ln(\lambda_1\lambda_2)}{\lambda_2}\label{P22},
\end{align}
where $\lambda_1$ and $\lambda_2$ are stretches in 1- and 2-directions, respectively. 
Equation \eqref{P11} can be solved for given values of $\lambda_2$, and equation \eqref{P22} can then be used to obtain the corresponding value of $P_{22}$.
The numerical results match the analytical results (Fig. \ref{uniaxial_schematic}(b)).

\subsection{Heat Transfer and Solidification Problem}
\begin{figure}[H]
  \centering
\includegraphics[scale=0.65]{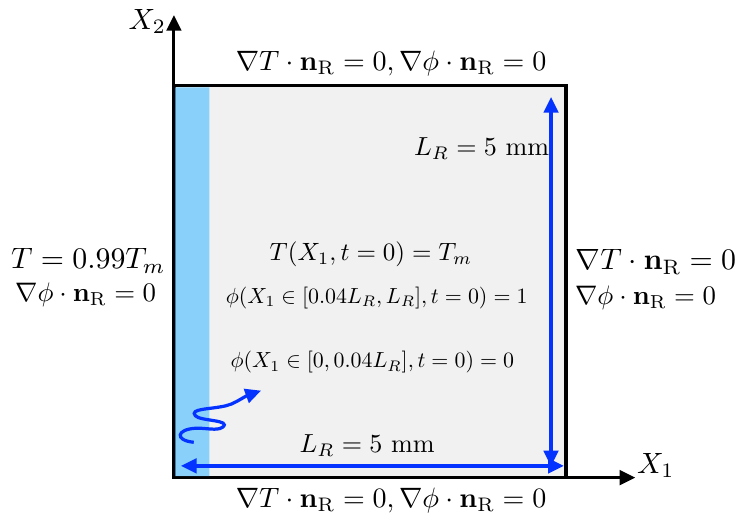}
\caption{Schematic of problem to study planar freezing front propagation. The domain was discretized natively in \texttt{FEniCS} using 160,000 structured quadrilateral elements.}\label{stefan_schematic}
 \end{figure}

We apply the model to capture the propagation of a planar freezing front in an effort to validate the results with the classic Stefan problem (Fig. \ref{stefan_schematic}).
For a purely thermal problem (ie. no mechanical coupling), the evolution equations for temperature and freezing phase field are
\begin{equation}
	C_v\dot{T} = K_{\rm R}\nabla^2 T  - \frac{T}{T_m}Lp'(\phi)\dot{\phi},
\end{equation}
and
\begin{equation}
	\dot{\phi} = -M\left(f_{\rm 0}g'(\phi) + L\frac{T_m-T}{T_m}p'(\phi) -\beta \nabla^2\phi\right).
\end{equation}
These equations are solved on a square domain of size $L_R \times L_R$ for the model parameters listed in Table \ref{tab:caseI-dim}.

\begin{table}[h]
  \centering
  \begin{minipage}{\textwidth}
    \centering
    \caption{Parameters for Solidification Validation Problem}\label{tab:caseI-dim}
    \begin{tabular}{|ll||ll|}
      \hline\hline
      Parameter  & Value  & Parameter  & Value \\
      \hline
      $C_v$ & $1.71\times 10^6$\ J/(kg K) \cite{campagnoli2021experimental, guntur2013temperature}& $f_0$ & $180$ J/m$^3$\\
      $K_{\rm R}$ & $0.5$ W/(mK)\cite{bianchi2022thermophysical} & $g(\phi)$ & $ \phi^2(1-\phi)^2$ \\
      $M$ & $\{10^{-6}, 10^{-5}, 10^{-4}\}$ m$^3$/Js & $p(\phi)$& $\phi^3(6\phi^2-15\phi +10)$\\
      $T_{m}$ & $273$ K & $\beta$ & $10^{-3}$ J/m\\
      $L$ & $1.4\times 10^8$ J/m$^3$ \cite{choi2008quantitative}&  $L_{R}$ & 5 mm\\
      $T(\bfX, t=0)$ & $T_m$ & $T(\text{left boundary})$ & $0.99 \ T_m$\\
      $\phi(X_1 \in [0.04, L_R], t=0)$ & $1$ &$\Delta X_1= \Delta X_2$ & 0.0125 mm\\
      \hline\hline
    \end{tabular}
  \end{minipage}  \hfill
\end{table}

\begin{figure}[h!]
  \centering
  \begin{subcaptionbox}{\label{fig_sol:sub2}}[0.32\textwidth]
    {\includegraphics[scale=0.23]{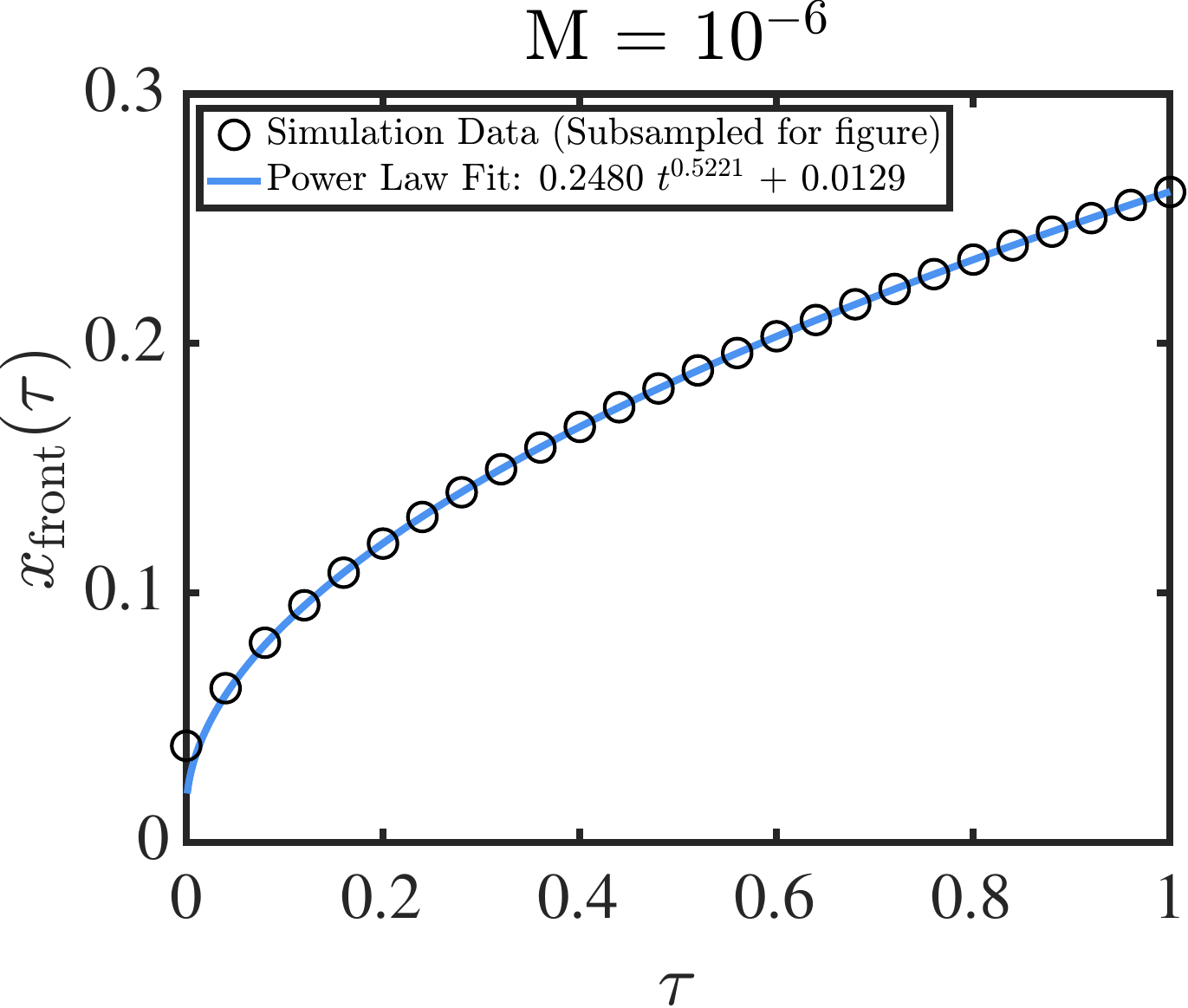}}
  \end{subcaptionbox}
  \begin{subcaptionbox}{\label{fig_sol:sub3}}[0.32\textwidth]
    {\includegraphics[scale=0.23]{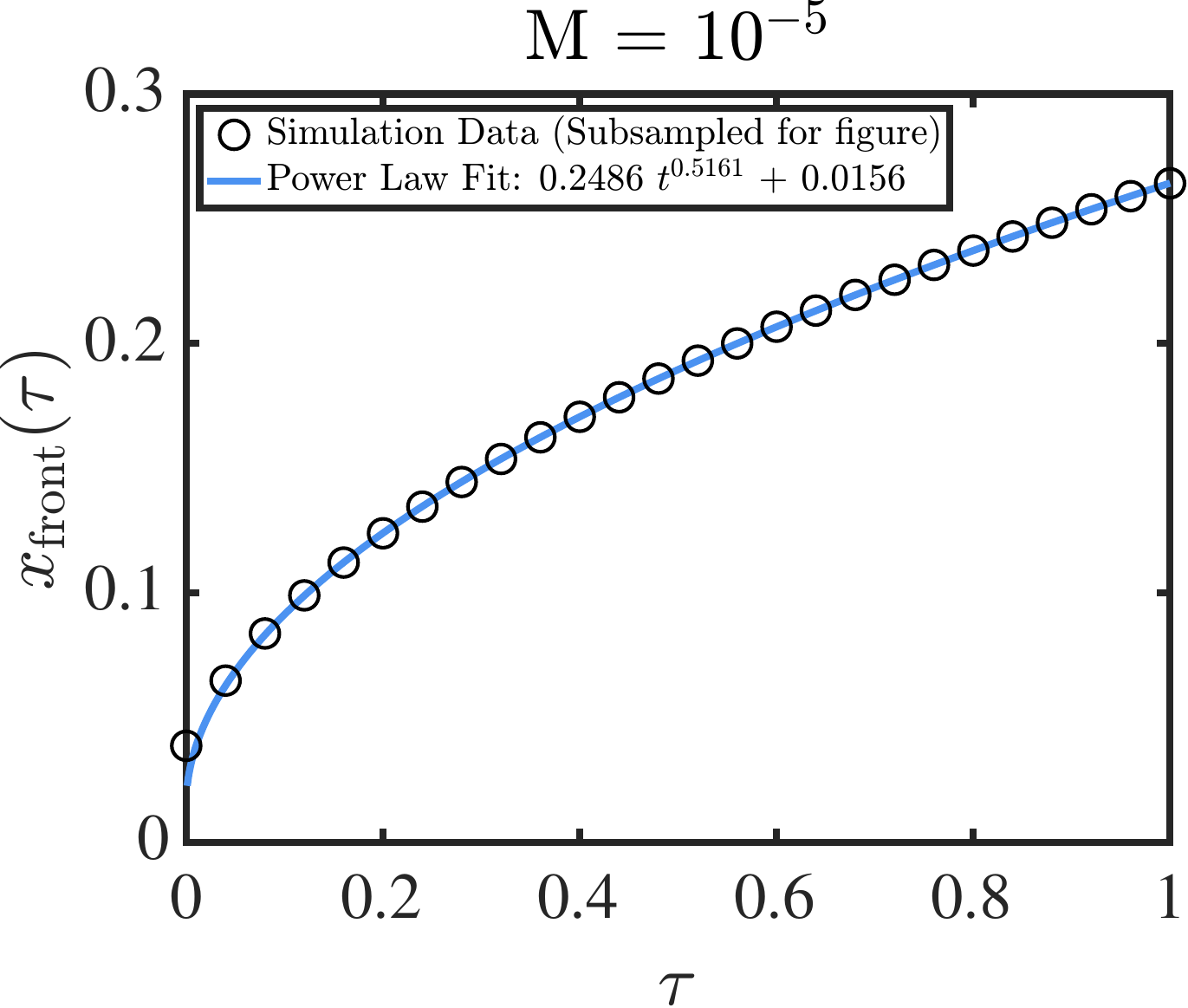}}
  \end{subcaptionbox}
  \begin{subcaptionbox}{\label{fig_sol:sub4}}[0.32\textwidth]
    {\includegraphics[scale=0.23]{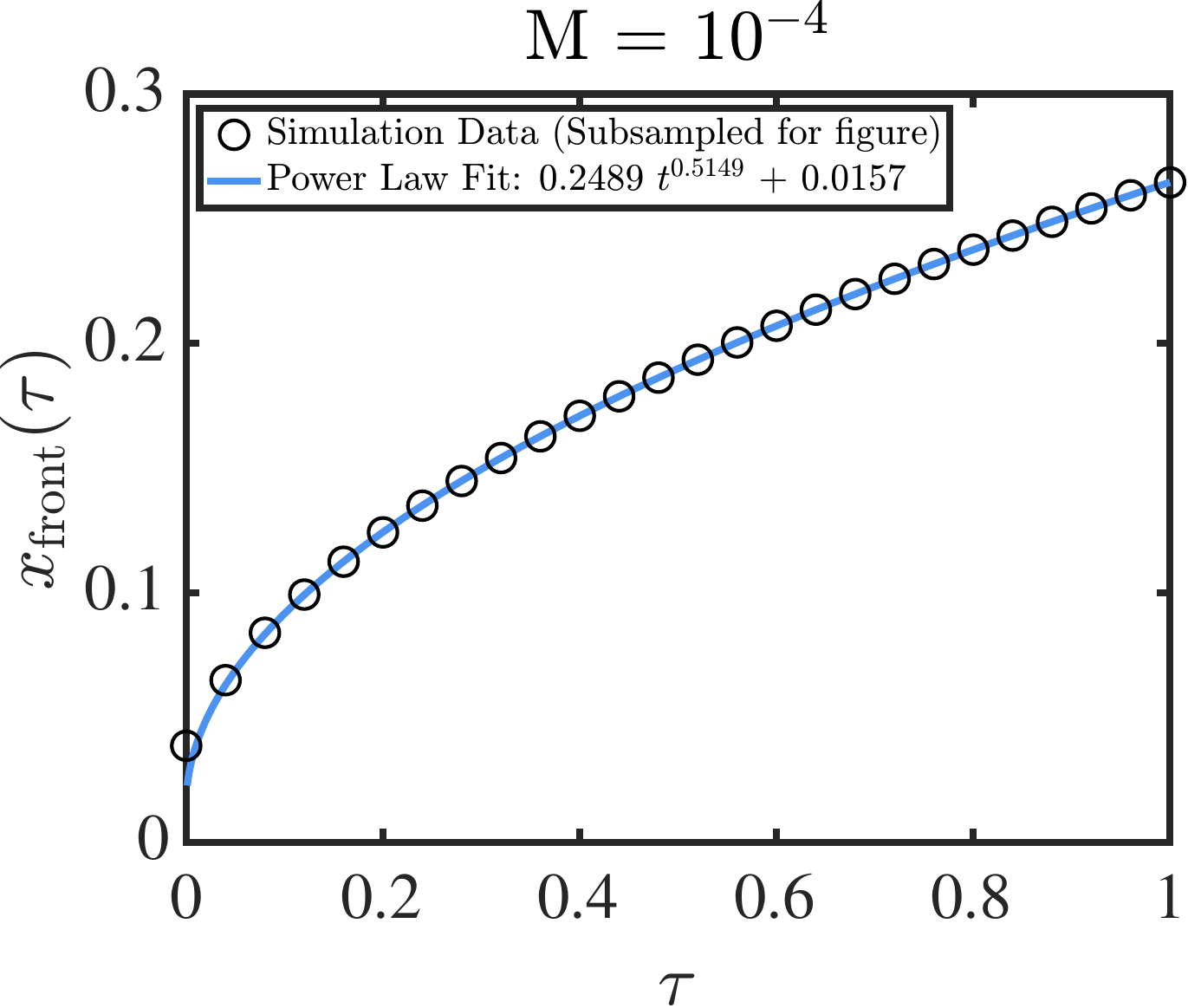}}
  \end{subcaptionbox}
  \caption{Location of the freezing front as a function of time for different values of the mobility parameter, $M$. Here, $\tau = t/t_R$, where $t_R = C_v L_R^2/K_R = 85.5$ s is a reference timescale.}
  \label{fig:three}
\end{figure}

Figure \ref{fig:three} shows the location of the freezing front (defined as the $X_1$-coordinate where $\phi = 0.5$) as a function of time $\tau = t/t_R$ for three different values of the mobility parameter $M$.
As the mobility increases, the results from the simulation approach the sharp interface Stefan's solution where the location of the front in an infinite, 1-D setting is proportional to $t^{1/2}$.
These results validate the thermal/phase change behavior of the framework and its \texttt{FEniCS} implementation.

\subsection{Damage Problem: Single Edge Notch Tension Test (SENT)}
\begin{figure}[h!]
  \centering
  \begin{subcaptionbox}{\label{fig_sol:sub1}}[0.45\textwidth]
    {\includegraphics[scale=0.7]{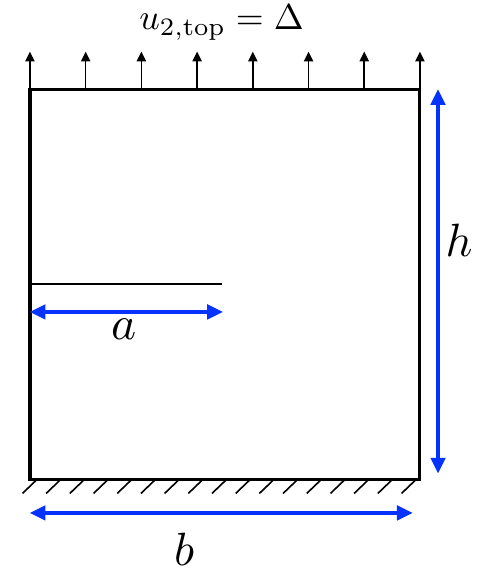}}
  \end{subcaptionbox}
  \begin{subcaptionbox}{\label{fig_sol:sub2}}[0.45\textwidth]
    {\includegraphics[scale=0.29]{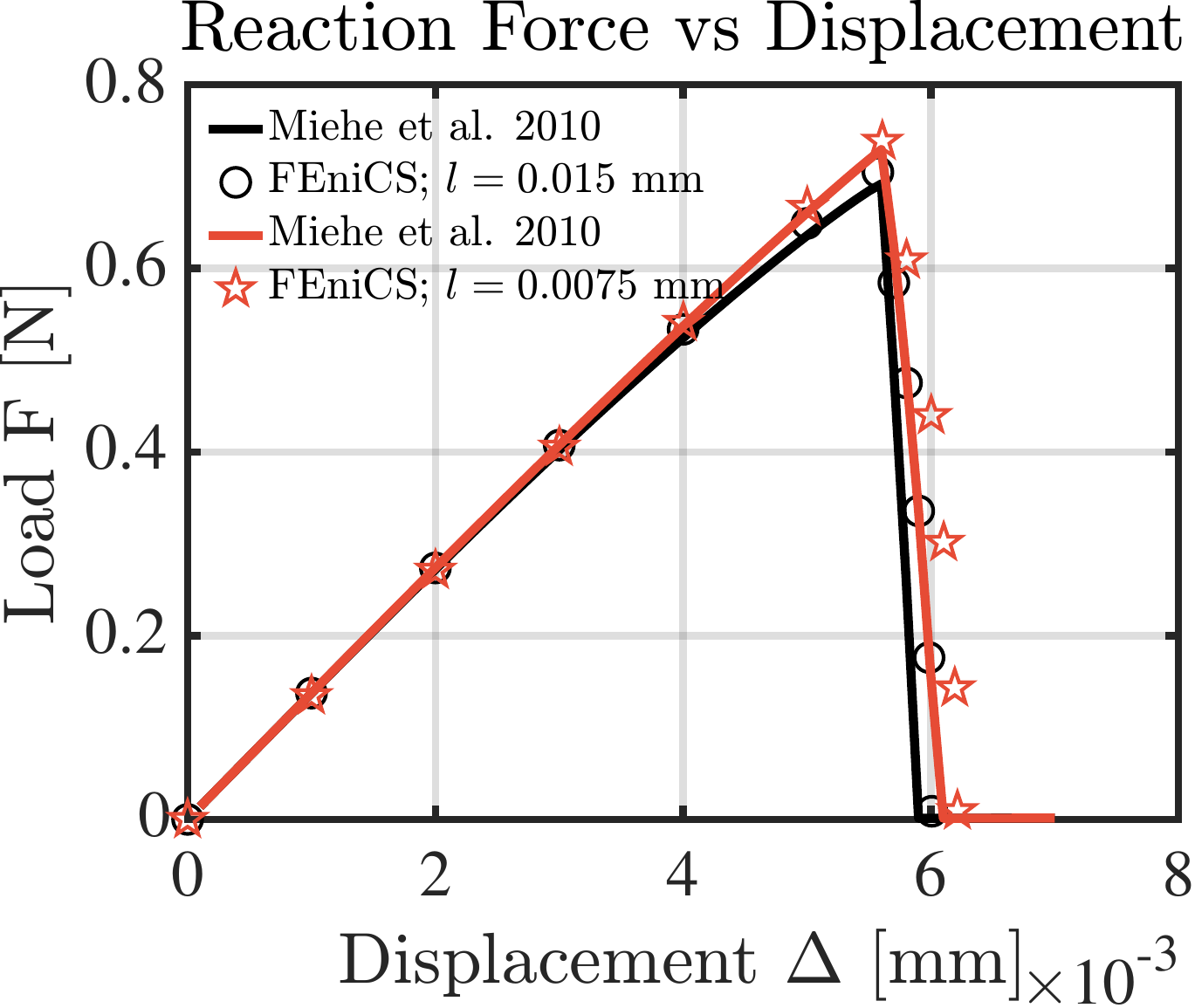}}
  \end{subcaptionbox}
   \caption{(a) Schematic of the SENT test. The mesh, prepared in Gmsh, contains 581,328 triangular elements. (b) Load-displacement curve from the validation test compared with results from Miehe et al. \cite{miehe2010phase}. Black color denotes the results for damage length parameter $l = 0.015$ mm and red color denotes the results for $l=0.0075$ mm.}
  \label{SENT_results}
\end{figure}

We validate the damage component of the model by applying it to a single edge notch tension test on a sample made of linear elastic material.
The setting is isothermal and the phase field is set to $\phi=1$.
Thus, only the mechanical part of the model is active.
In the linear elastic setting, there is no distinction between First Piola Kirchoff stress and Cauchy stress. 
These are given by
\begin{equation}
	\bfsigma = \frac{E}{(1+\nu)(1-2 \nu)}\left[ (1-2\nu)\bfepsilon + \nu  \rm{(tr{\bfepsilon})\id}\right]
\end{equation}
where $\bfepsilon$ is the infinitesimal strain tensor, and $E$ and $\nu$ are the Young's Modulus and Poisson's Ratio of the material, respectively.
Figure \ref{SENT_results} shows the schematic of the test.
The sample is rectangular with width $b$ and height $2h$.
The crack has a length $a$. 
Values used for these parameters are reported in Table \ref{tab:damage_val}.

\begin{table}[h]
  \centering
  \begin{minipage}{\textwidth}
    \centering
    \caption{Parameters for SENT validation}\label{tab:damage_val}
    \begin{tabular}{|ll||ll|}
      \hline\hline
		Parameter  & Value  & Parameter & Value \\
      \hline
      $E$ & $210$ kN/mm$^2$ & $a$ & $0.5$ mm \\
      $G_c$ & $2.7\times 10^{-3}$ kN/mm &  $b$ & $1$ mm \\
      $\nu$ & $0.3$ & $h$ & $1$ mm\\
       \hline\hline
    \end{tabular}
  \end{minipage}  \hfill
\end{table}

\newpage
\section{Representative Cases} 
In this section, we apply the specialized framework from \S3 to four representative cases.
While the framework is dimension-agnostic, we will consider plane strain setting for these cases.
We will use the material properties listed in Table \ref{tab:rep_cases}, unless stated otherwise.
Where available, parameters such as the elastic moduli ($\mu$, $K$) and thermal properties ($C_v$, $K_R$) have been chosen to be representative of soft biological tissues (e.g., liver).
The value of $G_c$ for brittle frozen tissues was not readily available, thus, its value has been chosen for convenience to illustrate the mechanics.
The phase-field parameters (e.g., $f_0$, $\beta$, $M$) govern the phenomenological aspects of the model, such as the energy barrier for phase change, the diffuse interface width, and the kinetic mobility. As these are not typically available from direct physical measurement, they have been selected to ensure numerical stability, convergence, and computational efficiency.
For a direct, quantitative comparison with specific experimental data, these model-specific parameters would be calibrated to match the observed experimental length- and time-scales.
\begin{table}[h]
  \centering
  \begin{minipage}{\textwidth}
    \centering
    \caption{Parameters for Representative Cases}\label{tab:rep_cases}
    \begin{tabular}{|ll||ll|}
      \hline\hline
      Parameter  & Value  & Parameter  & Value \\
      \hline
      $C_v$ & $1.71\times 10^6$\ J/(kg K) \cite{campagnoli2021experimental}& $f_0$ & $1.44 \times 10^3$ J/m$^3$\\
      $K_{\rm R}$ & $1$ W/(mK) \cite{campagnoli2021experimental, guntur2013temperature} & $g(\phi)$ & $ \phi^2(1-\phi)^2$ \\
      $M$ & $10^{-6}$ m$^3$/Js & $p(\phi)$& $\phi^3(6\phi^2-15\phi +10)$\\
      $T_{m}$ & $273$ K & $\beta$ & $4\times10^{-3}$ J/m\\
      $L$ & $1.4\times 10^8$ J/m$^3$ \cite{choi2008quantitative}&  $L_{R}$ & 10 mm\\
      $\mu$ & $10$ kPa \cite{millonig2010liver, hudson2013inter} & $K$ & $10^3 \times \mu$\\
      $\epsilon_0$ & $0.03$ &$g_d(d)$ & $(1-d)^2$ \\
      $G_c$ & $1.5$ N/m &$w(d)$ & $d^2$\\
      $l$ & $0.5$ mm &$\Delta x$ & 0.05 mm\\
      $\alpha_{\rm water}$ & $50\times 10^{-6}/^\circ$C &$\alpha_{\rm ice}$ & $50\times 10^{-6}/^\circ$C \\
      \hline\hline
    \end{tabular}
  \end{minipage}  \hfill
\end{table}


\subsection{Case I: Unconstrained freezing of a square sample}
We consider an $L_R \times L_R$ sample with temperature prescribed as  $T_0 = 193$\ K ($-80^\circ$C) on all boundaries.
  The sample is mechanically unconstrained, ie. all boundaries are traction-free.
Freezing and damage phase fields are subject to natural boundary conditions on all boundaries.\\

\begin{figure}[H]
  \centering
\includegraphics[scale=0.7]{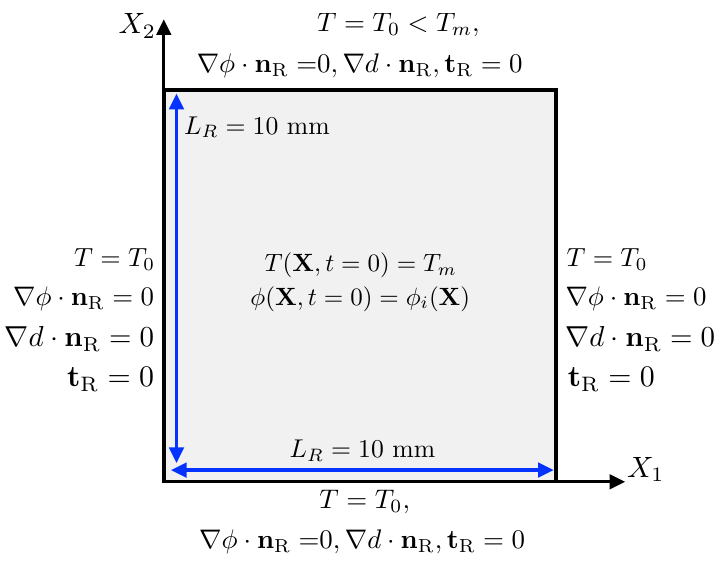}
\caption{Schematic of the unconstrained freezing setup. The domain was discretized natively in \texttt{FEniCS} using 40,000 structured quadrilateral elements.}\label{case1_schematic}
 \end{figure}

The initial temperature of the sample is set at $273$ K ($0^\circ$C), which is also taken to be the freezing temperature.
To initiate the freezing phase transformation in numerics, a thin layer near the boundaries, where $\phi$ transitions from a very low value $(\phi \sim 0)$ to unfrozen $(\phi = 1)$, is set as initial condition on all the boundaries, which is prescribed in accordance with the following form:
\begin{align}
    \phi_i(\mathbf{X}) &= \frac{1}{2} \left[ 1 + \tanh\left(\gamma \left( d(\mathbf{X}) - \delta \right) \right) \right], \label{eq:ic_c} \\
    \text{where} \quad d(\mathbf{X}) &= \min(X_1, L_R-X_1, X_2, L_R-X_2), \nonumber
\end{align}
$\gamma = 80/L_R$ controls the interface steepness, and $\delta = 0.02 L_R$ controls the interface width.
Figure \ref{case1_schematic} shows the schematic of the setup.\\

\begin{figure}[H]
  \centering
\includegraphics[scale=0.5]{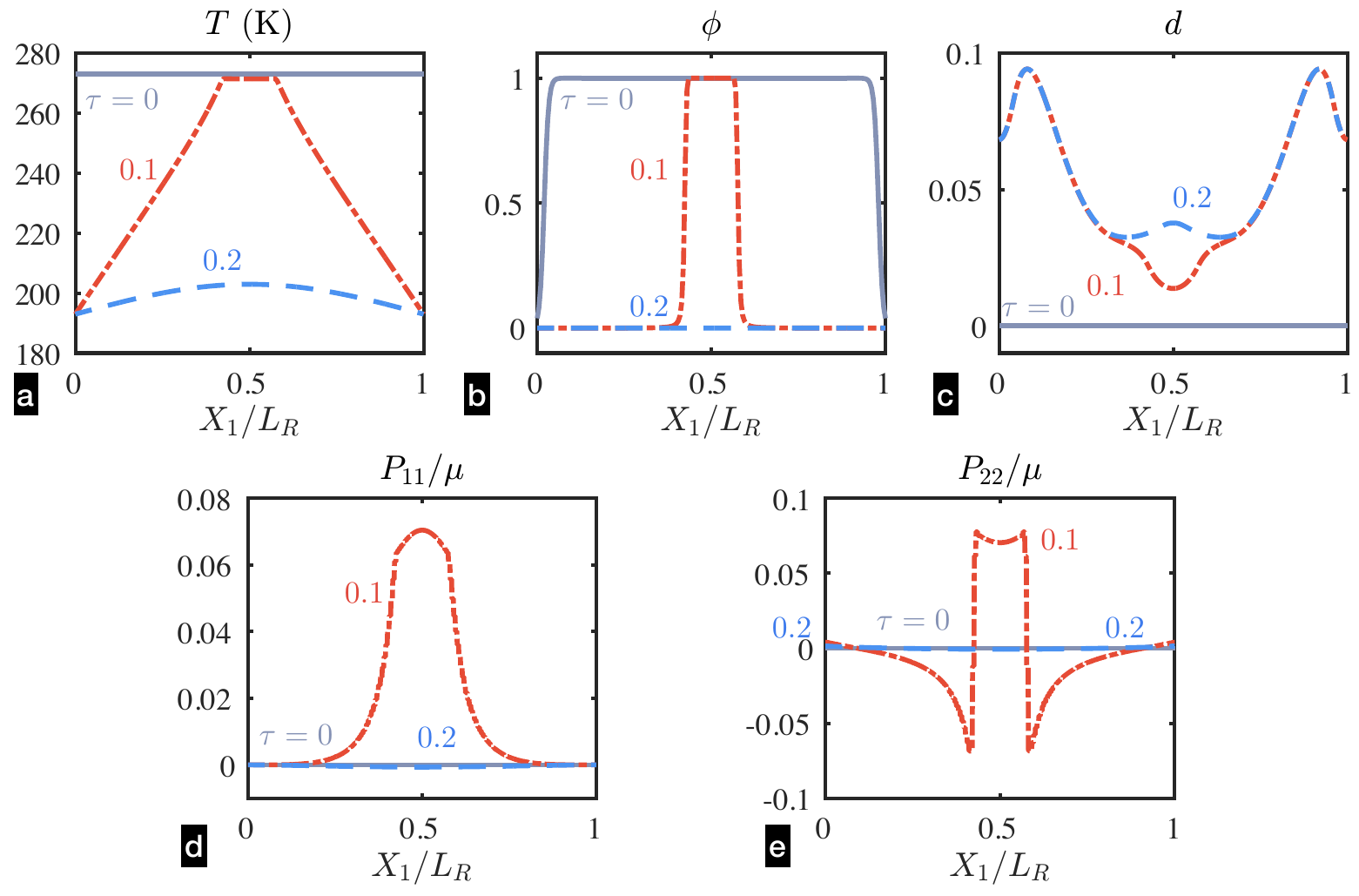}
\caption{Evolution of (a) temperature, (b) freezing phase field, (c) damage, (d) $P_{11}$, and (e) $P_{22}$ fields for the unconstrained freezing setup across $X_2/L_{R} = 0.5$ line. The fields are shown at three different instants of time: $\tau = t/t_R= 0, 0.1$, and $0.2$, where the reference time is $t_R =$ 171 s.}\label{case1_plots}
 \end{figure}

\begin{figure}[H]
\centering
\includegraphics[scale=0.5]{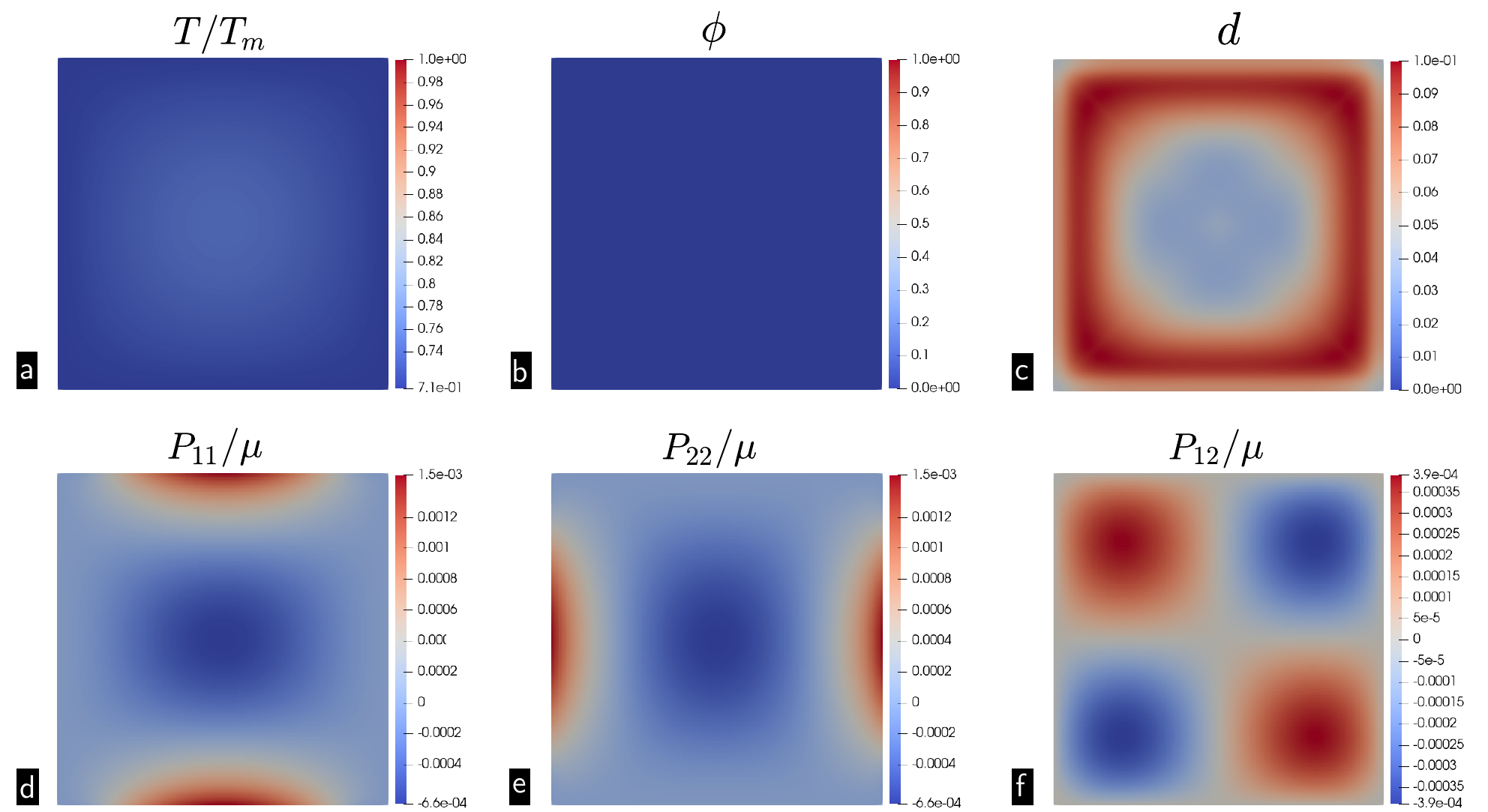}
\caption{Spatial distribution of (a) temperature, (b) freezing phase field, (c) damage, (d) $P_{11}$, (e) $P_{22}$, and (f) $P_{12}$ fields for the unconstrained freezing setup for Case I. The fields are shown at the end of the simulation at $\tau = t/t_R= 0.2$, where the reference time is $t_R =$ 171 s.}\label{case1_contour}
 \end{figure}

Figure \ref{case1_plots} shows the evolution of temperature, freezing phase field, damage, and stress fields along a plane through the center of the domain ($X_2/L_R = 0.5$), for three different instances of time: $\tau = t/t_R= 0, 0.1$, and $0.2$, where $t_R = C_v L_R^2/K_R$ is a reference timescale.
The simulations are carried out for a dimensional time of $34.2$s.
Contour plots in figure \ref{case1_contour} show the spatial evolution of various fields at the conclusion of the simulation ($\tau = 0.2$).
 The symmetry of the problem setup reflects in the evolution of the fields.\\
 
 As the sample starts to cool down, and undergo phase change, differential thermal and phase-change expansions across the sample induce stresses and damage; this can be seen in the dashed red plots for $\tau = 0.1$.
By the end of the simulation, the entire sample is nearly completely frozen and the temperature gradients are gentle.
Thus, the stresses in this unconstrained sample start to vanish, as is shown by the dashed blue plots in figure \ref{case1_plots}.
The damage generated during this evolution is, however, irreversible, and does not disappear when the entire sample has reached thermal and phase-change equilibrium.

\subsection{Case II: Constrained freezing of a square sample}
The thermal, freezing, and damage boundary conditions are the same as in case I, namely:
 we consider an $L_R \times L_R$ sample with temperature prescribed as $T_0=193$\ K ($-80^\circ$C) on all boundaries, and freezing and damage phase fields are subject to natural boundary conditions on all boundaries.
The sample is mechanically constrained: $u_2 = 0$ on the bottom  boundary ($X_2/L_R = 0$) and $u_2 = 0.02L_R$ on the top boundary ($X_2/L_R = 1$).\\

\begin{figure}[H]
  \centering
\includegraphics[scale=0.7]{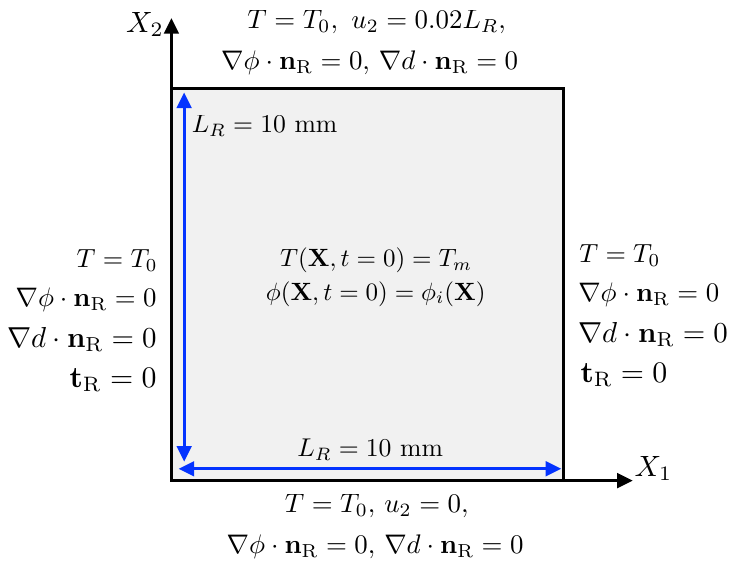}
\caption{Schematic of the constrained freezing setup. The domain was discretized natively in \texttt{FEniCS} using 40,000 structured quadrilateral elements.}\label{case2_schematic}
 \end{figure}

The initial conditions are exactly the same as that of case I, namely: the initial temperature of the sample is set at $273$ K ($0^\circ$C) and the freezing phase-field is initialized as $\phi(\bfX, t=0) = \phi_{i}(\bfX)$.
Figure \ref{case2_schematic} shows a schematic of the problem setup. \\

\begin{figure}[H]
  \centering
\includegraphics[scale=0.5]{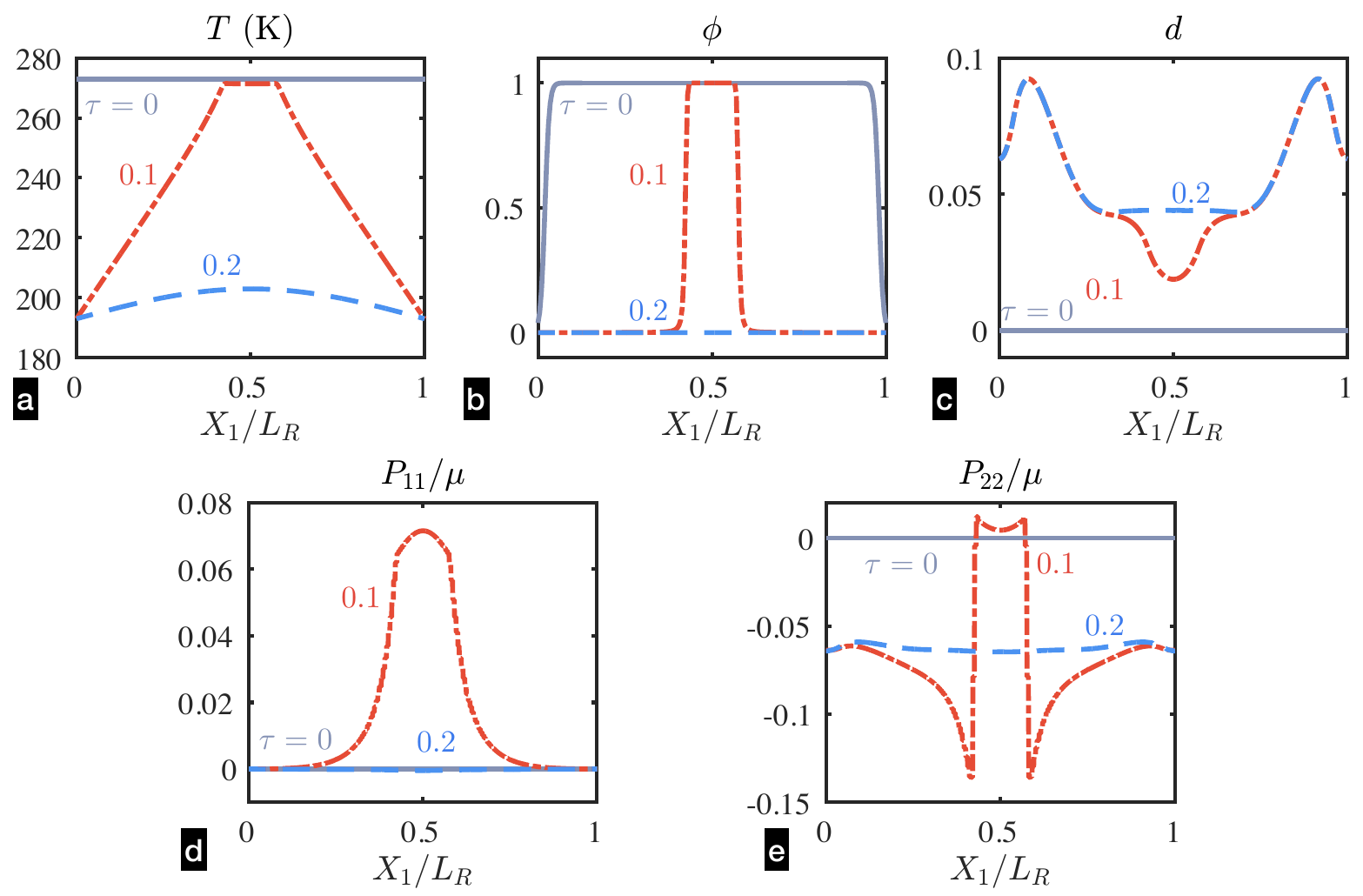}

\caption{Evolution of (a) temperature, (b) freezing phase field, (c) damage, (d) $P_{11}$, and (e) $P_{22}$ fields for the constrained freezing setup across $X_2/L_{R} = 0.5$ line. The fields are shown at three different instants of time: $\tau = t/t_R= 0, 0.1$, and $0.2$, where the reference time is $t_R =$ 171 s.}\label{case2_plots}
 \end{figure}

\begin{figure}[H]
\centering
\includegraphics[scale=0.5]{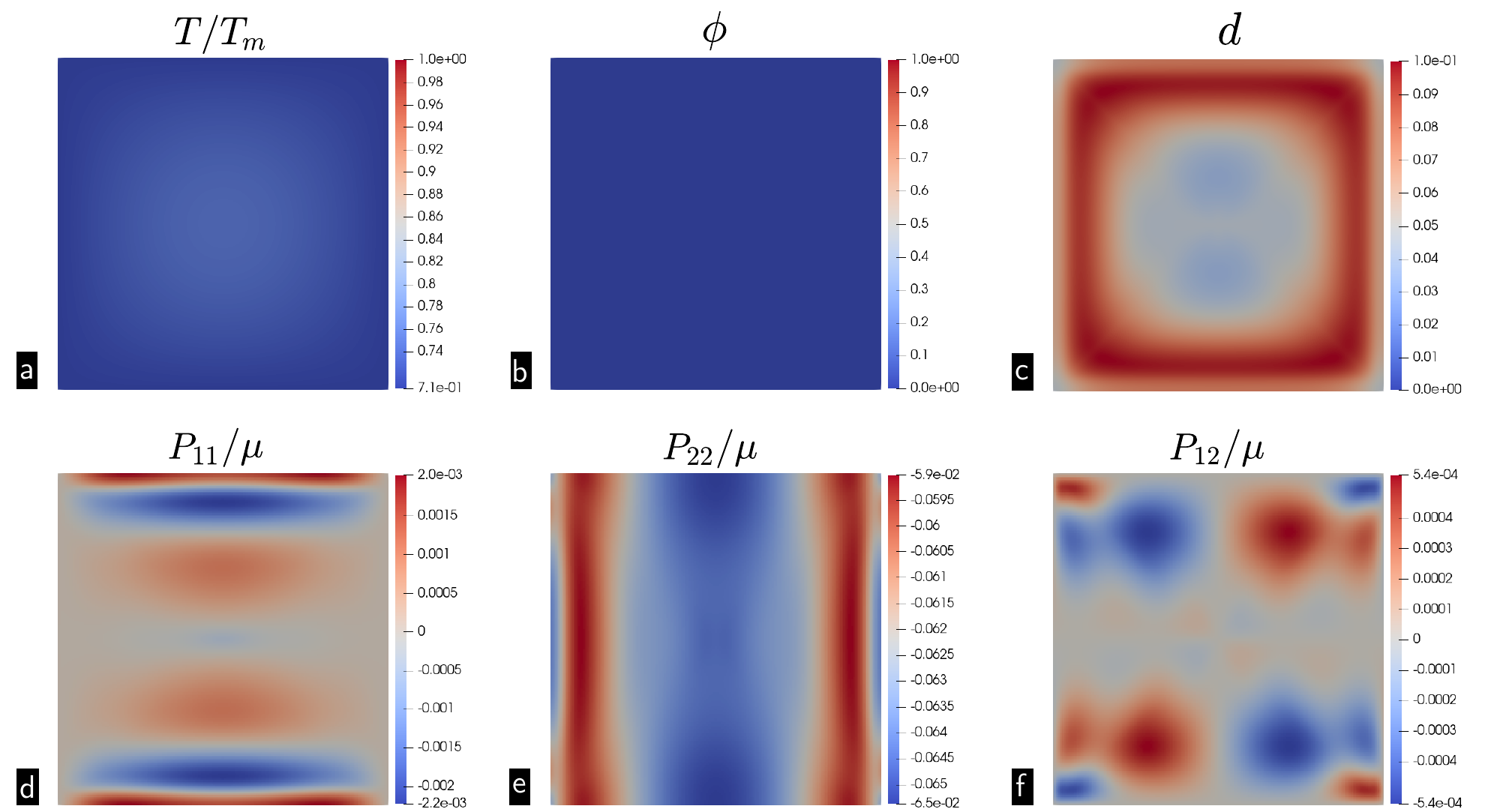}
\caption{Spatial distribution of (a) temperature, (b) freezing phase field, (c) damage, (d) $P_{11}$, (e) $P_{22}$, and (f) $P_{12}$ fields for the constrained freezing setup for Case II. The fields are shown at the end of the simulation at $\tau = t/t_R= 0.2$, where the reference time is $t_R =$ 171 s.}\label{case2_contour}
 \end{figure}

Figure \ref{case2_plots} shows the evolution of temperature, freezing phase field, damage, and stress fields along a plane through the center of the domain ($X_2/L_R = 0.5$), for three different instances of time: $\tau = t/t_R= 0, 0.1$, and $0.2$, where $t_R = C_v L_R^2/K_R$ is the reference timescale.
The simulations are carried out for a dimensional time of $34.2$s.
Contour plots in figure \ref{case2_contour} show the spatial evolution of various fields at the conclusion of the simulation ($\tau = 0.2$).\\

Similar to case I, differential thermal and phase-change expansions drive the stresses during the evolution of freezing in the sample.
The sample is nearly entirely frozen by the end of the simulation; in contrast to case I, the stresses do not vanish.
This is because the sample is mechanically constrained and the final state of stress reflects the imposed displacement boundary conditions in the 2-direction.
This is why $P_{22}$ is non-zero. 
Further, since the accumulated damage is non-uniform, the resulting $P_{22}$ is also non-uniform.
Notably, the peak damage in case I and case II are similar\footnote{\textit{Remark}: The location and magitude of peak damage is likely a function of the initial condition on $\phi$. Regardless, the model is able to capture any form of initial conditions, including ones that might be better physically motivated.}, possibly indicating that mechanical constraint applied is ``weak''.

\vspace{-20pt}

 \subsection{Case III: Freeze-thaw temperature conditions in square sample}
\begin{figure}[htbp]
\centering

\newlength{\panelht}
\setlength{\panelht}{0.29\textwidth} 

\begin{minipage}[t][\panelht][t]{0.49\textwidth}
  \centering
  \includegraphics[width=\linewidth]{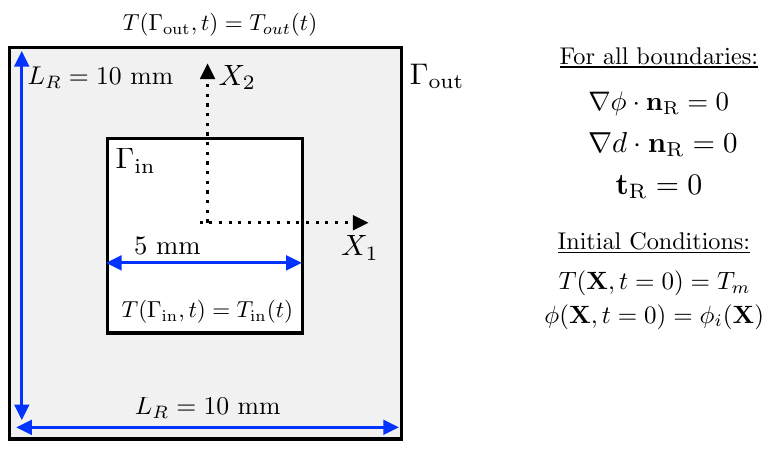}
  \vfill
  (a)
\end{minipage}
\hfill
\begin{minipage}[t][\panelht][t]{0.49\textwidth}
  \centering
  \resizebox{\linewidth}{!}{%
    \begin{tikzpicture}
      \definecolor{myCustomColor}{HTML}{3C5488}
      \begin{axis}[
          width=8cm, height=5cm, 
          xlabel={$t$}, ylabel={$\bar{T}(t)/T_m$},
          axis x line*=bottom, axis y line*=left,
          xmin=0, xmax=1.05,
          ymin=0.65, ymax=1.25,
          grid=major,
          scaled ticks=false,
          xtick={0, 0.25, 0.5, 0.75, 1.0},
          xticklabels={$0$, $\tfrac{P}{4}$, $\tfrac{P}{2}$, $\tfrac{3P}{4}$, $P$},
          ytick={0.7, 1.0, 1.2},
          yticklabels={$0.7$, $1.0$, $1.2$},
          ticklabel style={font=\small},
          label style={font=\small},
          legend style={font=\small, cells={anchor=west}, at={(0.03,0.97)}, anchor=north west}
      ]
        \addplot[very thick, red,  solid] coordinates {(0,1.0) (0.25,0.7)};
        \addlegendentry{Outer Boundary}
        \addplot[very thick, blue, solid, forget plot] coordinates {(0.25,0.7) (0.5,1.0)};
        \addplot[very thick, teal, solid, forget plot] coordinates {(0.5,1.0) (0.75,1.0)};
        \addplot[very thick, myCustomColor, solid, forget plot] coordinates {(0.75,1.0) (1.0,1.0)};

        \addplot[very thick, red,  dashed] coordinates {(0,1.0) (0.25,1.0)};
        \addlegendentry{Inner Boundary}
        \addplot[very thick, blue, dashed, forget plot] coordinates {(0.25,1.0) (0.5,1.0)};
        \addplot[very thick, teal, dashed, forget plot] coordinates {(0.5,1.0) (0.75,1.2)};
        \addplot[very thick, myCustomColor, dashed, forget plot] coordinates {(0.75,1.2) (1.0,1.2)};

        \draw[dashed, thick] (axis cs:0.25,0.65) -- (axis cs:0.25,1.25);
        \draw[dashed, thick] (axis cs:0.5, 0.65) -- (axis cs:0.5, 1.25);
        \draw[dashed, thick] (axis cs:0.75,0.65) -- (axis cs:0.75,1.25);

        \node[anchor=south west] at (axis cs:0,    1.0) {\scriptsize A};
        \node[anchor=north west] at (axis cs:0.25, 0.7) {\scriptsize B};
        \node[anchor=south west] at (axis cs:0.5,  1.0) {\scriptsize C};
        \node[anchor=south west] at (axis cs:0.75, 1.0) {\scriptsize D};
        \node[anchor=south west] at (axis cs:1.0,  1.0) {\scriptsize E};
      \end{axis}
    \end{tikzpicture}
  }
  \vfill
  (b)
\end{minipage}

\caption{(a) Schematic of the unconstrained freeze/thaw setup in a square annulus. The mesh, prepared in Gmsh, contains 17,694 triangular elements. (b) Programmed temperature evolution over one period \(P\). Points A--E mark transitions between phases. The fields are shown for $P = 8.55$ s.}\label{case3_schematic}

\end{figure}
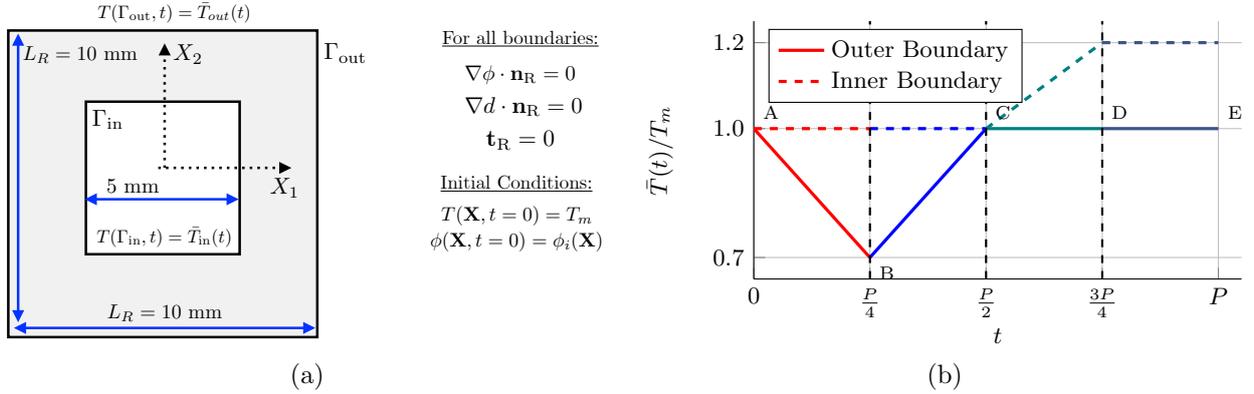

We consider a square annulus geometry in this case, as shown in Figure \ref{case3_schematic}(a).
Temperature is prescribed on all boundaries and is a function of time, given by $\bar{T}_{\rm in}(t)$ on the inner boundary and $\bar{T}_{\rm out}(t)$ on the outer boundary.
The functional forms for temperature boundary conditions are given by equations \eqref{eq:temperature_piecewise1} and \eqref{eq:temperature_piecewise2} for a time period $P$ and are visually represented in Figure \ref{case3_schematic}(b).
The prescribed temperature evolution is designed to show the behavior of the sample under freezing and subsequent thawing conditions, which are typical of cryopreservation protocols.
\begin{equation}\label{eq:temperature_piecewise1}
	\bar{T}_{\text{out}}(t)/T_m = 
\begin{cases} 
1.0 - (0.3) \cdot \frac{t}{P/4} & \text{if } 0 \le t \le \frac{P}{4} \\
0.7 + (0.3) \cdot \frac{t - P/4}{P/4} & \text{if } \frac{P}{4} < t \le \frac{P}{2} \\
1.0 & \text{if } \frac{P}{2} < t  \leq P
\end{cases}
\end{equation}

\begin{equation}\label{eq:temperature_piecewise2}
\bar{T}_{\text{in}}(t)/T_m = 
\begin{cases} 
1.0 & \text{if } 0 \le t \le \frac{P}{2} \\
1.0 + (0.2) \cdot \frac{t - P/2}{P/4} & \text{if } \frac{P}{2} < t \le \frac{3P}{4} \\
1.2 & \text{if } \frac{3P}{4} < t \leq P
\end{cases}
\end{equation}

The initial conditions are exactly the same as that of cases I and II, namely: the initial temperature of the sample is set at $273$ K ($0^\circ$C) and the freezing phase-field is initialized as $\phi(\bfX, t=0) = \phi_{i}(\bfX)$.\\

\begin{figure}[H]
  \centering
\includegraphics[scale=0.5]{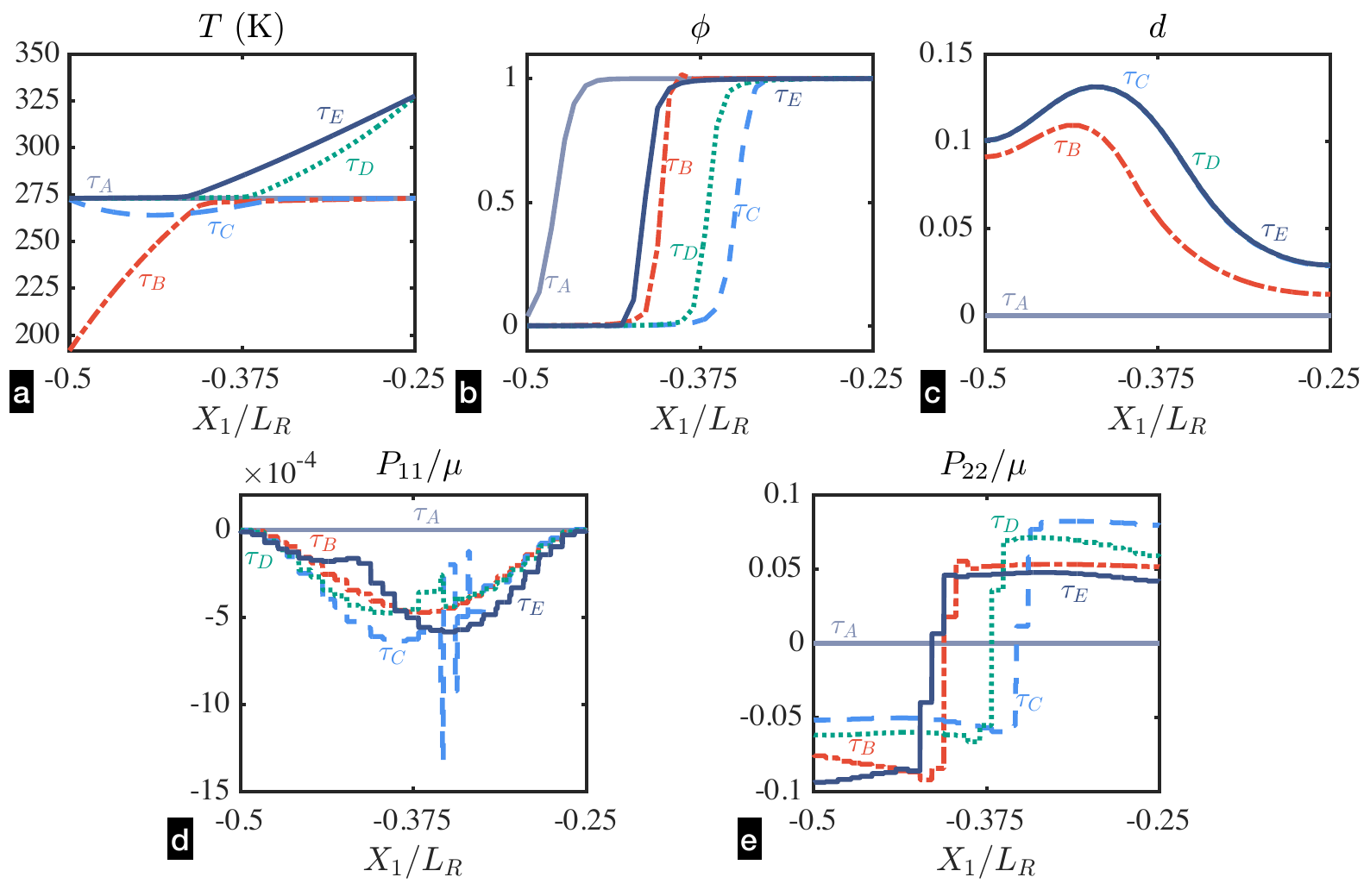}

\caption{Evolution of (a) temperature, (b) freezing phase field, (c) damage, (d) $P_{11}$, and (e) $P_{22}$ fields for the constrained heating setup across $X_2/L_{R} = 0.5$ line. The fields are shown at five different instants of time: $\tau := t/t_R, \tau_A= 0, \tau_B = 0.0125, \tau_C = 0.025, \tau_D = 0.0375$, and $\tau_E = 0.05$, where the reference time is $t_R =$ 171 s.}\label{case3_plots}
\end{figure}

\begin{figure}[H]
\centering
\includegraphics[scale=0.45]{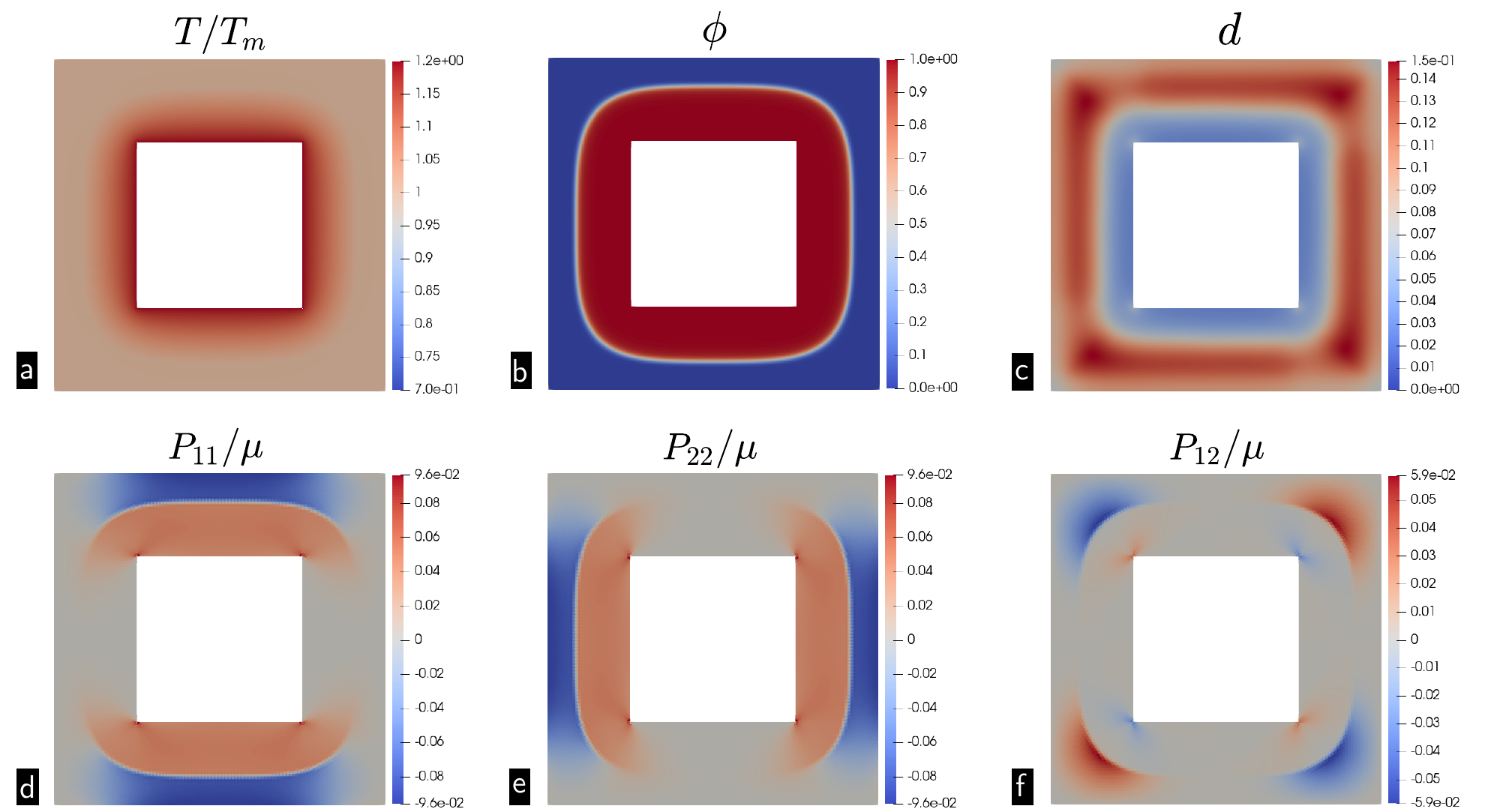}
\caption{Spatial distribution of (a) temperature, (b) freezing phase field, (c) damage, (d) $P_{11}$, (e) $P_{22}$, and (f) $P_{12}$ fields for the unconstrained freezing setup for Case III. The fields are shown at the end of the simulation at $\tau = t/t_R= 0.05$, where the reference time is $t_R =$ 171 s.}\label{case3_contour}
 \end{figure}

For this particular study, we use $\beta = 8\times 10^{-3}$ J/m.
Figure \ref{case3_plots} shows snapshots of fields at different instances of time along the line $X_2/L_R = 0$.
During the cooling and then warming phase back to freezing temperature on the outer boundary (from A-C), the freezing front proceeds inwards from the boundaries (Fig. \ref{case3_plots}(b); $\tau_A \rightarrow \tau_B \rightarrow \tau_C$), which leads to development of stresses and induces damage.
The temperature is then held constant at $T_m$ on the outer boundary; the inner boundary temperature is raised slowly to $T = 1.2 T_m$ (C-D) and then held constant (D-E).
This triggers thawing of the sample, and the freezing front starts to reverse back towards the outer boundary, as can be seen in the movement of the freezing from $\tau_C \rightarrow \tau_D \rightarrow \tau_E$. 
Notably, damage is induced in the freezing process but not in the thawing process in this case.
This happens because of the following reason: the transformation strain from freezing is the dominant contributor to the stress.
As the freezing front moves inwards, it causes a spike in the strain energy density across the front, which drives damage.
During the thawing process, the strain energy density across the front stays below the previously  experienced peak strain energy and thus does not induce additional damage (see Figure \ref{case3_damageEnergyDensity}).
Just like in Case I, the boundaries of the sample are mechanically unconstrained; a stronger constraint of all boundaries being fixed rapidly induces a much higher damage in the sample (in results not included in the present study).
The dominant stress component along the $X_2/L_R = 0$ line is $P_{22}$; $P_{11}$ remains much smaller during the entire process.
The spatial distribution of fields at the end of the simulation is presented in figure \ref{case3_contour}.
\vspace{-1em}

\vspace{-20pt}
 
\subsection{Case IV: Fixed temperature and displacement boundary conditions in an irregular shape}
In this case, we consider an irregular shape to showcase the broad applicability of the model.
A fixed temperature $T_0= 193$\ K ($-80^\circ$C) is prescribed on the entire boundary.
The displacements are constrained such that the entire boundary is fixed.
Natural boundary conditions apply to damage and freezing phase fields. \\

The entire domain is initially at the freezing temperature of $T_m = 273$ K.
To initiate freezing, a thin strip around the boundary of the domain that smoothly evolves from $\phi=0$ to $\phi=1$ is prescribed as the initial condition.
\vspace{-10pt}
\begin{figure}[H]
\centering
\includegraphics[width = 0.4\textwidth]{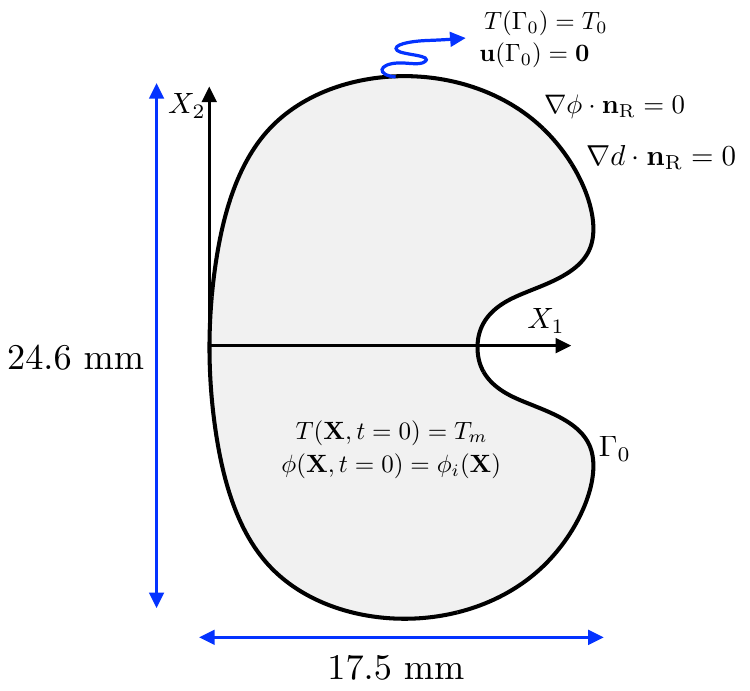}
\caption{Schematic of the kidney with fixed temperature and displacement boundary conditions. The mesh, prepared in Gmsh, contains 316,782 triangular elements.}\label{case4_schematic}
 \end{figure}
For this particular study, we use $G_c = 1$ N/m, $\mu = 1$ kPa, and $\beta = 8\times 10^{-3}$ J/m. 
Other parameters remain the same as in Table \ref{tab:rep_cases}.
Figure \ref{case4_plots} shows the evolution of temperature, freezing phase field, damage fields, and stresses along a plane through the center of the domain ($X_2/L_R = 0$), for three different instances of time: $\tau = t/t_R= 0$, $0.00356$, and $0.00712$,  where $t_R = C_v L_R^2/K_R$ is a reference timescale.
The simulations are carried out for a dimensional time of $\sim 1.2$ s.
Contour plots in figure \ref{case4_contour} show the spatial evolution of various fields at the conclusion of the simulation ($\tau = 0.00712$).

We find that the mechanical constraint imposed in this problem is ``stronger'', in contrast to Case II, since the damage starts to develop early on in the entire sample (see also Appendix A for more detailed time evolution contour plots).
\vspace{-1em}
\begin{figure}[H]
  \centering
\includegraphics[scale=0.5]{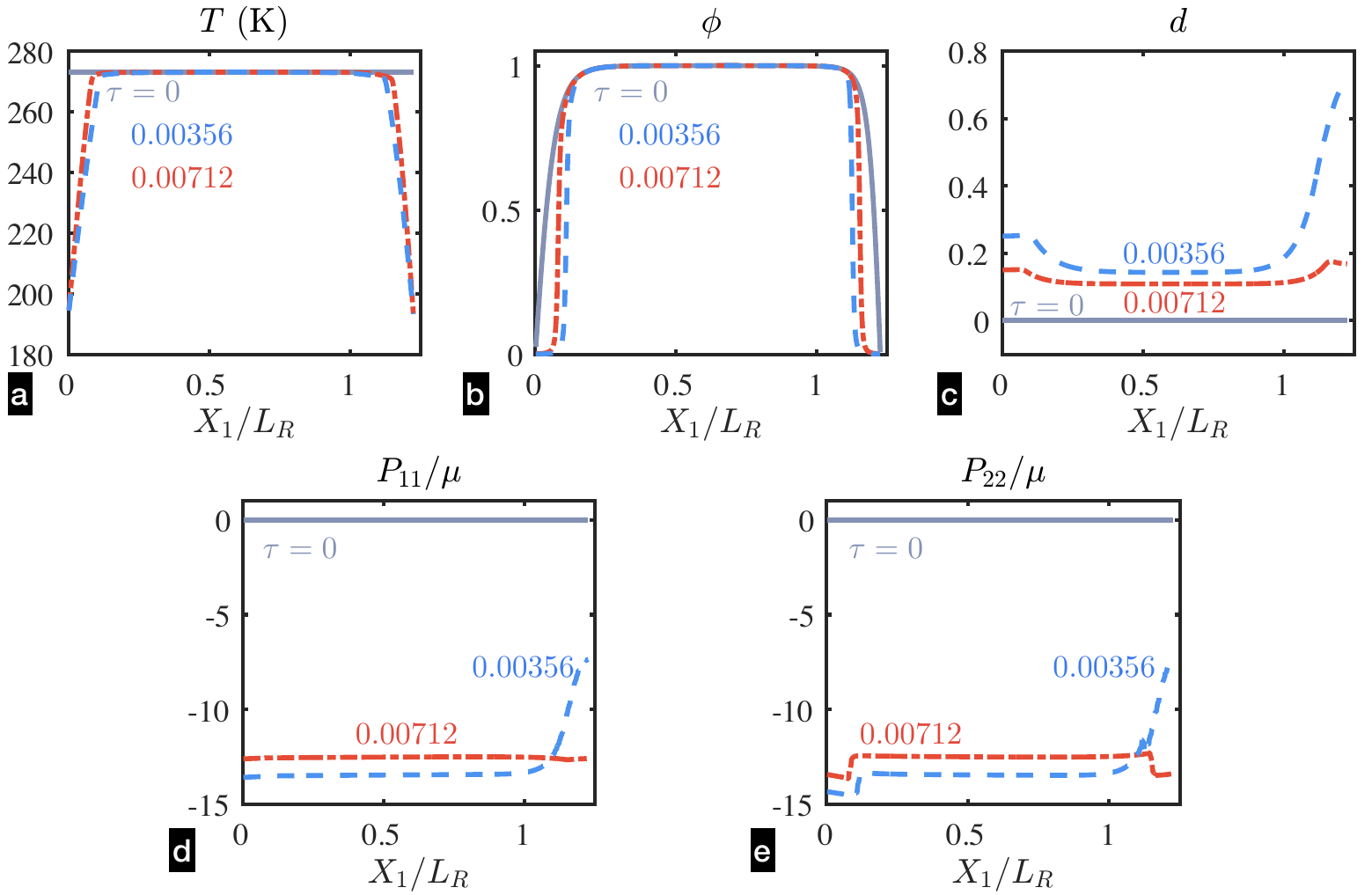}

\caption{Evolution of (a) temperature, (b) freezing phase field, (c) damage, (d) $P_{11}$, and (e) $P_{22}$ fields for the constrained heating setup across $X_2/L_{R} = 0$ line. The fields are shown at three different instants of time: $\tau = t/t_R= 0$, $0.00356$, and $0.00712$, where the reference time is $t_R =$ 171 s.}\label{case4_plots}
 \end{figure}
 The damage appears to severely localize near the curved region along the center of the domain.
 The curvature tends to concentrate stresses, which in turn drive damage.
 In general, regions of stress concentrations are more prone to damage.
 
\begin{figure}[h]
  \centering
\includegraphics[scale=0.5]{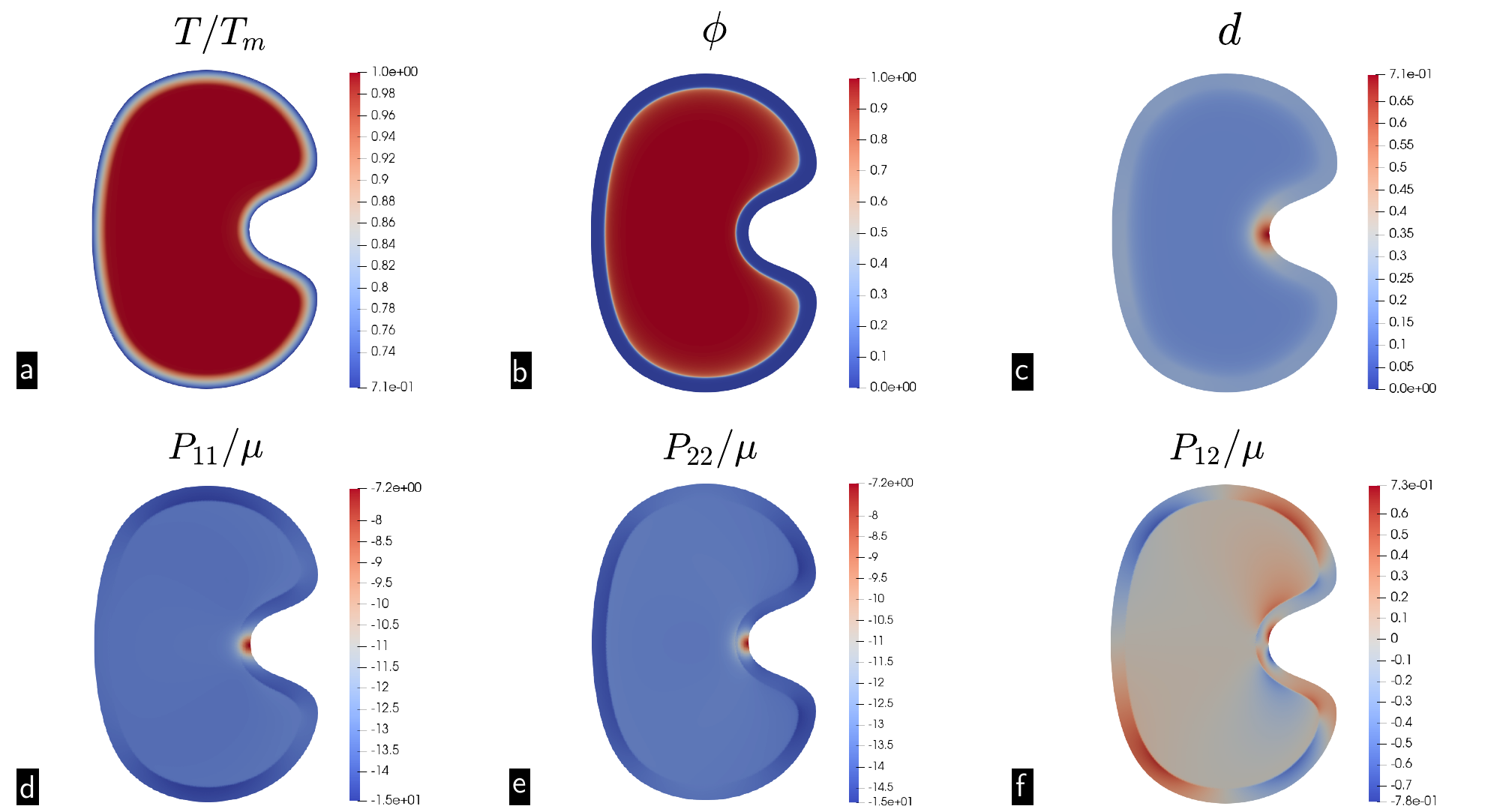}
\caption{Spatial distribution of (a) temperature, (b) freezing phase field, (c) damage, (d) $P_{11}$, (e) $P_{22}$, and (f) $P_{12}$ fields for the constrained heating setup for Case IV. The fields are shown at the end of the simulation at $\tau = t/t_R= 0.00712$, where the reference time is $t_R =$ 171 s. Due to the strong mechanical constraint of fixed boundary on the entire domain, the damage grows rapidly in this system.}\label{case4_contour}
 \end{figure}
\section{Discussion, Conclusion, and Future Work}\label{conclusions}
We have developed a thermodynamically consistent, thermo-mechanically coupled finite deformation model for capturing the freezing-induced damage in soft materials.
The developed framework is by nature modular in the way it accounts for various sources of energy contributions, and therefore, easily extensible to include other physics, if necessary.
The representative cases showcase the capability of the model to capture a diverse range of thermal and mechanical scenarios.
Results also show that mechanical constraints can significantly affect the stresses experienced by the sample and the resulting damage.
In case II, the mechanical constraint of applied displacement in the 2-direction is ``weaker'' than the constraint applied in case IV, where displacement on all boundaries is zero.
The latter, thus, shows significantly higher damage starting much earlier in the process.\\

Current work is a first step in the direction of developing a comprehensive and predictive understanding of the cryopreservation mechanics.
While this work successfully integrates multiple physical processes into a comprehensive framework, it is worth nothing that there are important open problems which need to be addressed in future works on this topic. Below we list some such opportunities for investigation:
\begin{itemize}
	\item As touched upon in the introduction, the model focuses on the development of bulk thermo-mechanical stresses developed in the second stage of the freezing process, and  does not account for internal transport of water.
Furthermore, the model does not have an explicit freezing nucleation mechanism built in.
Addressing these two phenomena in the modeling can widen the applicability of the model to early stages of the freezing process as well.
	\item The choice of damage model is an area with significant scope for more exploration. While the current framework provides a baseline for damage evolution, future works will focus on integrating more sophisticated models. The current formulation can be extended to include rate-dependent damage, which is likely important as tissues are subjected to protocols involving varying rates of cooling and thawing.
Future works will integrate other damage models proposed recently in the mechanics community that could more accurately simulate the initiation and propagation of damage across a range of samples, from pristine materials to those with existing cracks.
	\item Lastly, the present work focuses on a simplified abstraction of tissue as a nonlinear elastic material.
Future works may focus on a more nuanced description of the tissue microstructure including cells and extracellular matrix. This may be achieved in simplified continuum-scale models through inclusion of spatial variation of material properties, anisotropy of elastic and thermal properties, and the incorporation of viscoelasticity, which accounts for the time-dependent mechanical response of soft biological materials.
\end{itemize}
The thermodynamically consistent approach presented here and the resulting framework provide a robust foundation to advance modeling efforts in future, including in all of the above directions.

\section*{Acknowledgements}
We would like to acknowledge helpful discussions with Dr. Chockalingam Senthilnathan (Purdue University).

\section*{Declaration of Interest}
Declarations of interest: none
\appendix

\section{Detailed Time Evolution of Fields for Representative Cases} 

\begin{figure}[H]
	\centering
	\includegraphics[width = 1\textwidth]{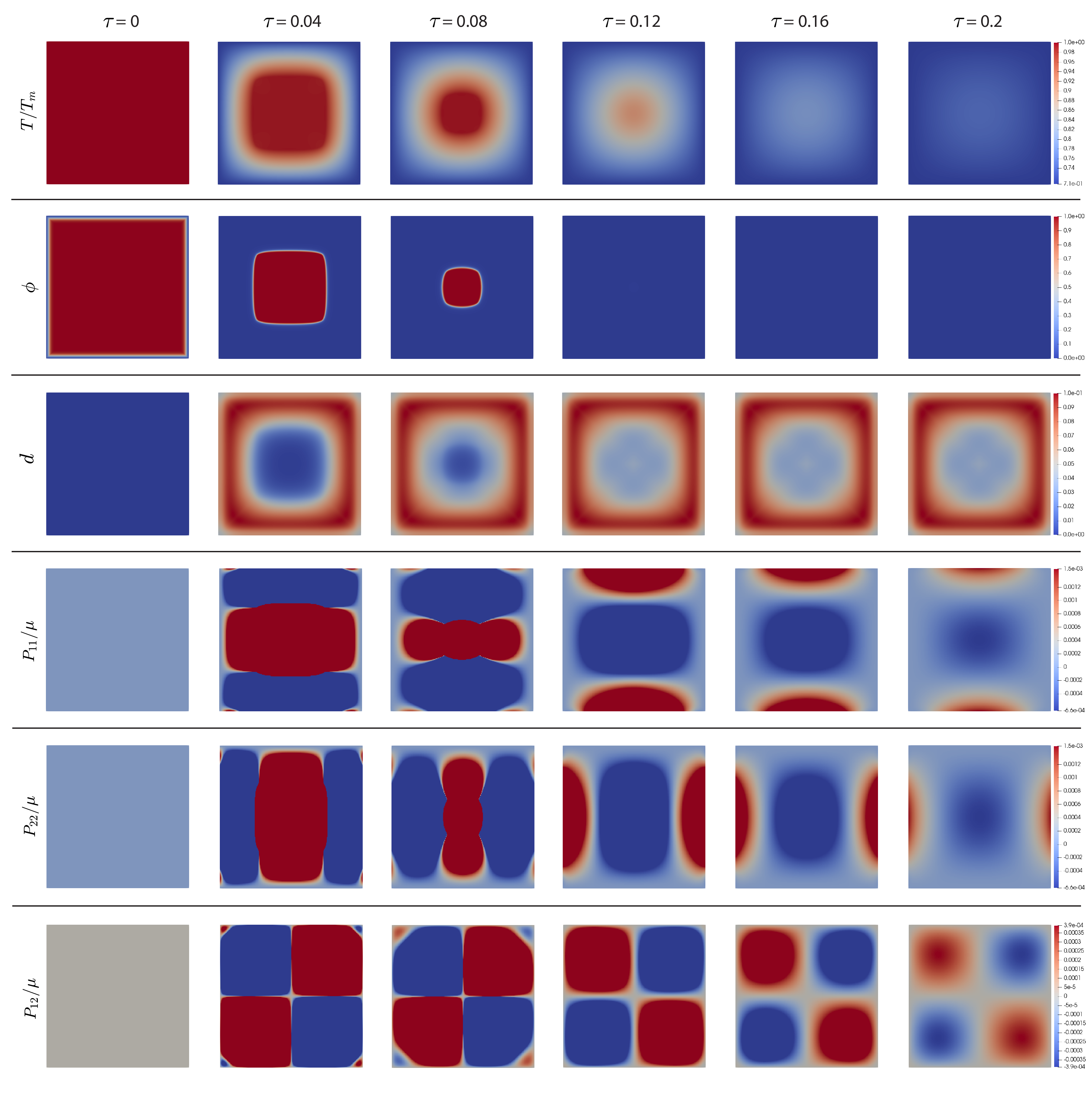}
	\caption{Contour plots showing the time evolution of various fields for unconstrained freezing case.}\label{case1_evolution}
 \end{figure}

\begin{figure}[H]
	\centering
	\includegraphics[width = 1\textwidth]{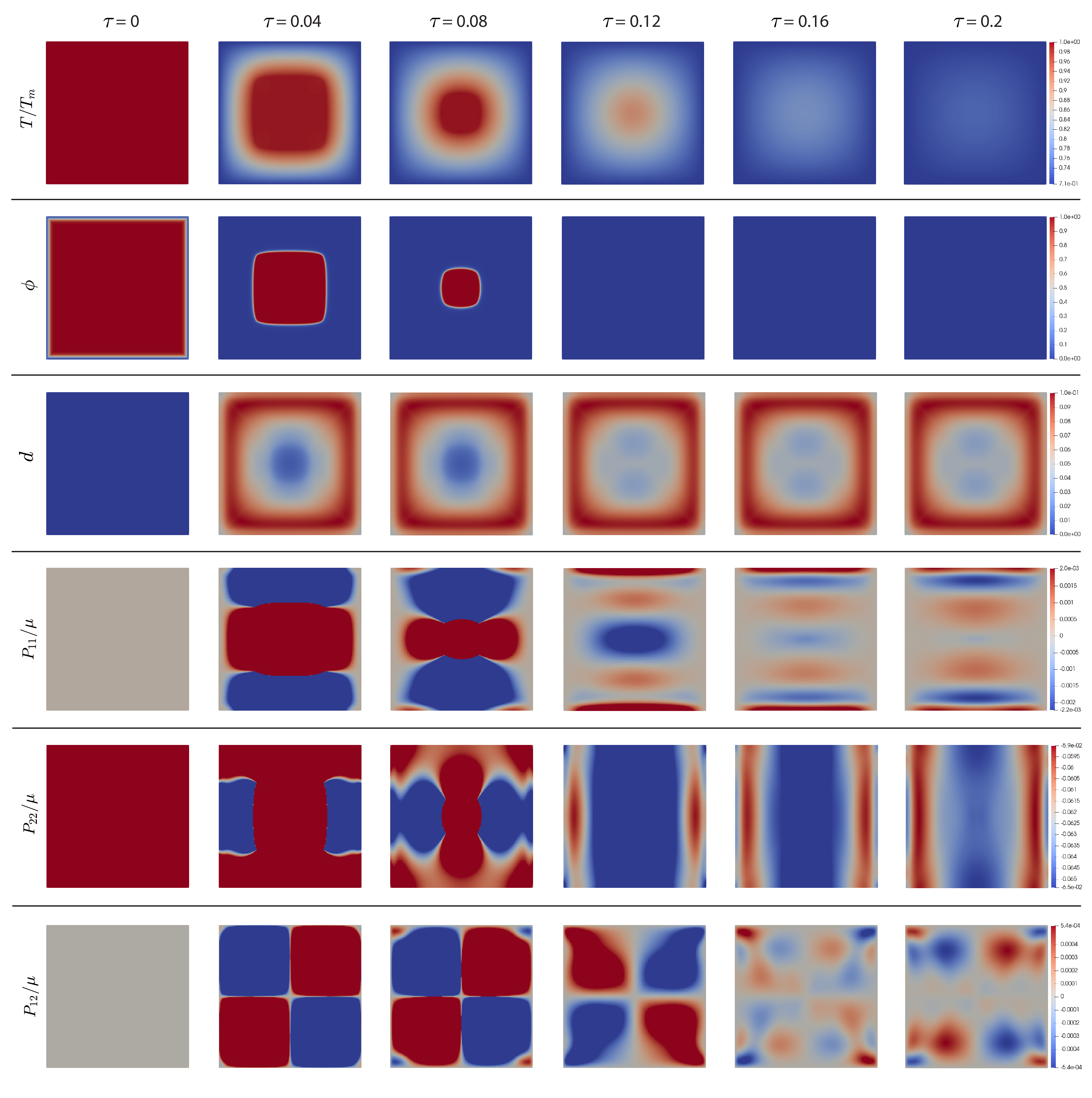}
	\caption{Contour plots showing the time evolution of various fields for constrained freezing case.}\label{case2_evolution}
 \end{figure}

\begin{figure}[H]
	\centering
	\includegraphics[width = 1\textwidth]{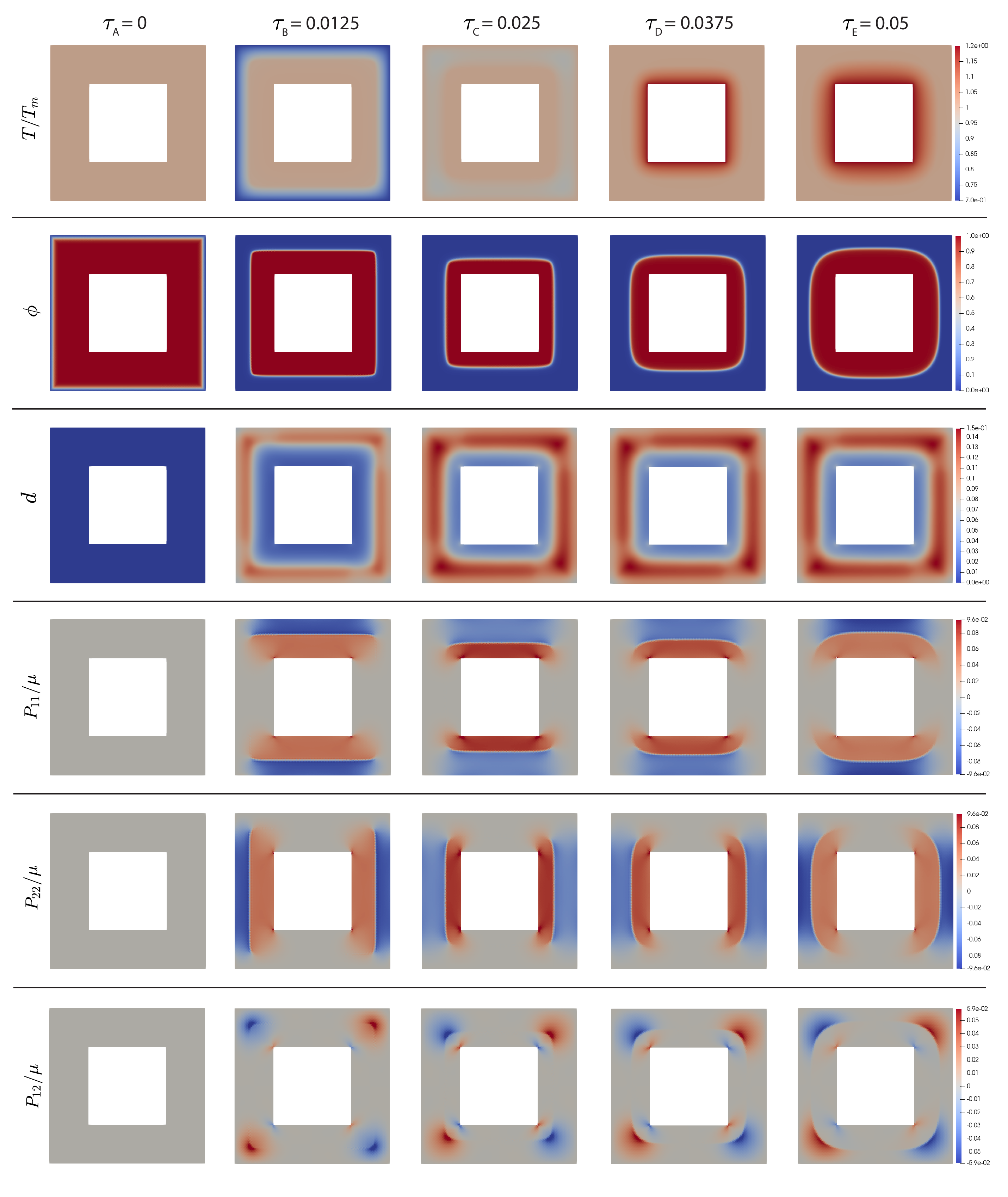}
	\caption{Contour plots showing the time evolution of various fields for unconstrained freezing in case of a square annulus.}\label{case3_evolution}
 \end{figure}

\begin{figure}[H]
	\centering
	\includegraphics[width = 0.6\textwidth]{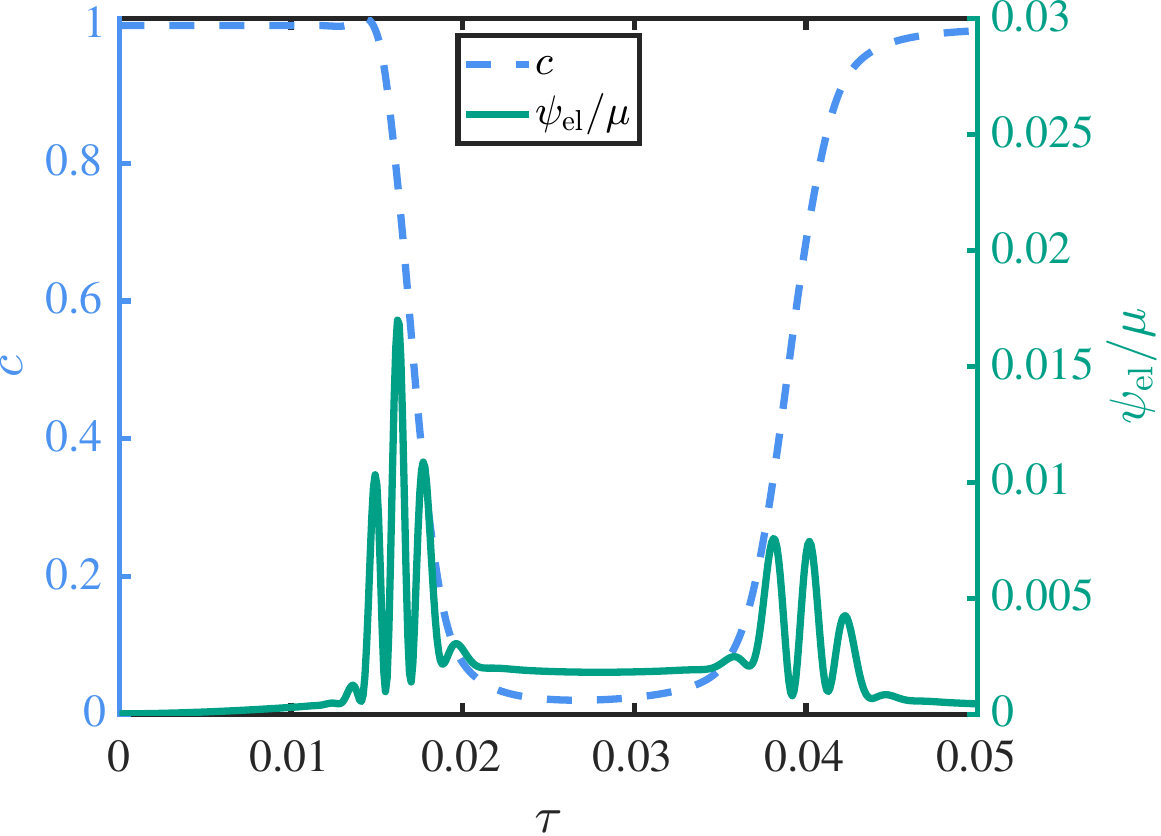}
	\caption{Plot of elastic energy density and freezing phase field at $(X_1/L_R, X_2/L_R) = (-0.375, 0)$ as a function of time. The maximum strain energy is experienced during the inward movement of the freezing front, which dictates the extent of damage. The thawing process does not incur additional damage because the peak strain energy density remains lower than during freezing.}\label{case3_damageEnergyDensity}
 \end{figure}

\begin{figure}[H]
	\centering
	\includegraphics[width = 0.6\textwidth]{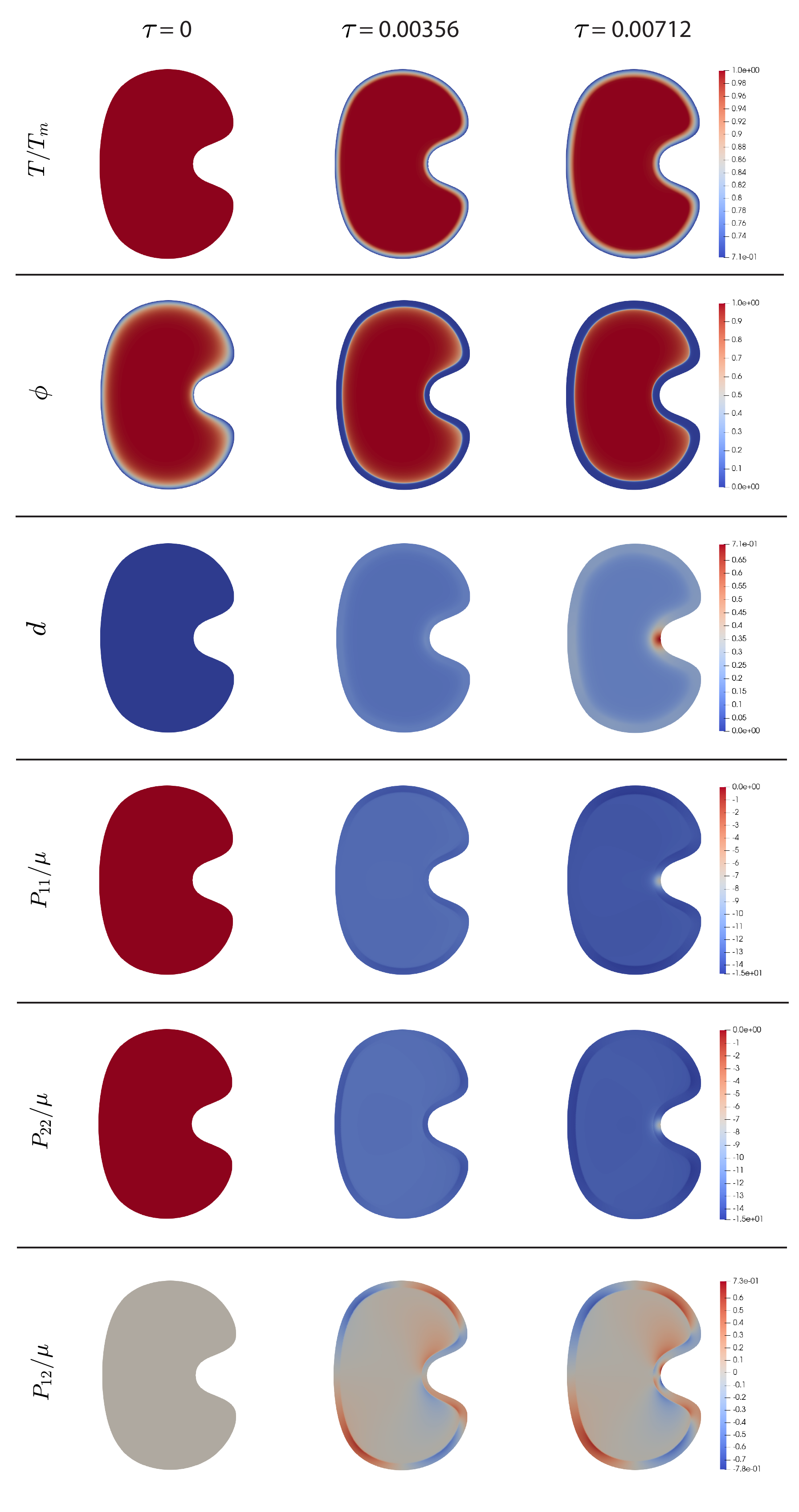}
	\caption{Contour plots showing the time evolution of various fields for constrained freezing case in an irregular shape. }\label{case4_evolution}
 \end{figure}
\bibliography{references}

\end{document}

%% file: MACROS_BK4.tex
\newcommand{\bfb}{{\bf b}}\newcommand{\bfB}{{\bf B}}%
\newcommand{\bfC}{{\bf C}}%
\newcommand{\bfF}{{\bf F}}%
\newcommand{\bfg}{{\bf g}}%
\newcommand{\bfl}{{\bf l}}%
\newcommand{\bfn}{{\bf n}}%
\newcommand{\bfo}{{\bf o}}%
\newcommand{\bfP}{{\bf P}}%
\newcommand{\bfq}{{\bf q}}\newcommand{\bfQ}{{\bf Q}}%
\newcommand{\bft}{{\bf t}}%
\newcommand{\bfv}{{\bf v}}%
\newcommand{\bfx}{{\bf x}}\newcommand{\bfX}{{\bf X}}%
\newcommand{\bfy}{{\bf y}}%
\newcommand{\bfbeta}{\boldsymbol{\beta}}%
\newcommand{\bfepsilon}{\boldsymbol{\epsilon}}%
\newcommand{\bfzeta}{\boldsymbol{\zeta}}%
\newcommand{\bfxi}{\boldsymbol{\xi}}%
\newcommand{\bfpi}{\boldsymbol{\pi}}%
\newcommand{\bfsigma}{\boldsymbol{\sigma}}%
\newcommand{\bfchi}{\boldsymbol{\chi}}%
\newfont{\tenbfit}{cmmib10}%
\newfont{\svnbfit}{cmmib8}%
\newfont{\tenbfsl}{cmbxti10}
\newfont{\mmit}{cmmi10}
\newfont{\smit}{cmmi9}
\newfont{\bfMit}{cmmi5}
\newfont{\tenbbb}{msbm10}%
\newfont{\svnbbb}{msbm8}%
\newfont{\tenssit}{cmssqi8 at 10pt}%
\newfont{\svnssit}{cmssqi8 at 7pt}%
\newfont{\gothic}{eufm10}%
\newfont{\sgothic}{eufm7}%

\newcommand{\pards}[2]{\mbox{$\dfrac{\partial #1}{\partial {#2 }}$}}

\newcommand{\Blj}{\mbox{$\Big[\kern-0.275em\Big[$}}
\newcommand{\Brj}{\mbox{$\Big]\kern-0.275em\Big]$}}

\newcommand{\id}{{\bf 1}}
\newcommand{\X}{\bfX}
\newcommand{\x}{\bfx}

\newcommand{\F}{\bfF}

\newcommand{\C}{\bfC}

\newcommand{\n}{\bfn}
\newcommand{\dev}{\text{dev}}
\newcommand{\Ce}{\bfC^e}
\newcommand{\Cedot}{\dot{\bfC}^e}